# Extreme Localization via a Singular Edge Mode in an Asymmetric Linkage System

Myeonggyun Joo[1], Dongwook Kwak[1], Rajesh Chaunsali[2], Panayotis G. Kevrekidis[3,4], and Jinkyu Yang[1]

[1]Department of Mechanical Engineering, Seoul National University, Seoul 08826, Republic of Korea
[2]Department of Aerospace Engineering, Indian Institute of Science, Bangalore 560012, India
[3]Department of Mathematics and Statistics, University of Massachusetts, Amherst, Massachusetts 01003-4515, USA
[4]Department of Physics, University of Massachusetts, Amherst, Massachusetts 01003, USA

## Abstract

Mechanical localized modes generally retain exponentially decaying tails, whereas confinement to a single structural unit is difficult to achieve without resonators, flat-band lattices or topological design. To address this need, an asymmetric linkage system is introduced in which geometry tunes a boundary mode from finite localization to its extreme limit. Each cell comprises two unequal rigid bars connected by a torsional spring and is described by one generalized coordinate. Linearization and coordinate transformation yield a tridiagonal model and closed-form expressions for the edge mode frequency and spatial decay ratio. The analysis identifies the configurations supporting a singular edge mode at which its decay ratio vanishes, leaving only the boundary cell active in the linear regime. Simulations show near-complete localization under finite-amplitude excitation, and experiments reproduce the configuration-dependent transition from bulk transmission to boundary confinement. These results establish a minimal and analytically tunable platform for mechanical localization.

## 1 Introduction

Mechanical metamaterials provide a powerful platform for controlling elastic energy flow through geometry rather than material characteristics alone. By designing unit cells and controlling their interactions, these systems exhibit unusual responses such as wave attenuation[1–3], nonlinear wave propagation[4–7], and directional wave transport[8–10], opening new possibilities for vibration isolation[11–13], impact protection[14, 15], and reconfigurable structures[16, 17]. For example, locally resonant and phononic metamaterials can selectively filter, steer, or attenuate waves through engineered dispersion relations [18–20] and lead to remarkable wave propagation properties. In addition, architected materials have enabled diverse nonlinear energy-transport phenomena, including amplitude-dependent wave dynamics and solitary-wave propagation [21–26].

Within this broader effort to manipulate dynamic energy transport, significant attention has been directed toward localizing mechanical energy at targeted regions. Topological metamaterials have provided a rich framework for this purpose, where boundary, interface, or defect modes can concentrate dynamic motion and energy in spatially confined regions while retaining robustness against perturbations.[27–30] Related localization phenomena have also been extensively studied in nonlinear optical lattices [31, 32]. More broadly, such localized modes provide a strategy for spatially programmable energy concentration, allowing mechanical response to be selectively amplified, isolated, or triggered at desired locations within an architected system.

Existing platforms for energy transfer and localization employ a broad range of mechanisms, including architected or reconfigurable structures [16, 17, 33], compliant continuum-based or flexible architectures [21–23], multistable energy landscapes [17, 21, 22], local resonators [11–13, 18], and topologically engineered lattices [27–30]. These approaches have enabled remarkable control over mechanical energy, although their governing mechanisms and design procedures are often tailored to the specific architecture and may involve multiple geometric, material, or coupling parameters. Moreover, many localized modes retain a finite spatial decay length. By contrast, confinement within a single or a few unit cells has typically been achieved through locally resonant topological designs or compact localized states arising from flat-band engineering [34–37]. Such extreme localization, in which energy penetration into the surrounding structure is nearly or completely suppressed, is particularly attractive for vibration isolation, impact protection, and spatially selective energy control. Accordingly, there remains a need for minimal mechanical systems that admit simple closed-form design relations directly connecting unit-cell geometry to the existence, frequency, and spatial decay of localized modes, while enabling localization to be tuned to its extreme limit.

Linkage-based systems can offer a promising solution for achieving such extreme localization within an analytically tractable framework. Since their motion is primarily governed by kinematic constraints of rigid links and joints, linkage structures can often be reduced to low-dimensional generalized coordinates, making them far more analytically accessible than continuum-based or high-dimensional deformable architectures. At the same time, linkages can serve as fundamental building blocks for a wide range of mechanical metamaterials. Previous studies[33, 38, 39] have shown that linkage-based unit cells can provide clearly defined deformation paths and programmable effective properties, suggesting that linkages can function not only as simple mechanical models but also as design platforms for more complex metamaterial architectures. Furthermore, bar-linkage systems have been showcased to possess remarkable features of their wave structures, including multiple types of moving solitary waves [28, 40], or the presence of neutral (Goldstone-type) translation modes, as well as of internal breathing modes in their linearization spectra [41].

Here, we introduce an asymmetric linkage system that supports extreme energy localization through an analytically tunable edge mode. Each unit cell consists of two unequal rigid bars connected by a torsional spring. Owing to the kinematic constraint of the linkage, the configuration of each cell can be described by a single generalized coordinate, enabling a compact formulation of the equation of motion. The resulting chain exhibits strongly configuration-dependent wave behavior, allowing either bulk energy transmission or edge-localized responses. Through modal analysis, we derive closed-form expressions that directly relate the unit-cell geometry to the edge-mode frequency and spatial decay, and show that they accurately capture the numerical simulation results. Remarkably, when the spatial decay coefficient expressed in terms of the geometric parameters vanishes, the localization length of the edge mode converges to zero; that is, only the first unit cell possesses a nonzero amplitude, while all subsequent cells exhibit exactly zero amplitude in the ideal linear system. Following the terminology used for singular topological edge states in locally resonant metamaterials [34], we refer to this limiting, compactly supported boundary state as a *singular edge mode*, where "singular" denotes the vanishing localization length. Although the previously reported and present states share this limiting localization behavior, their underlying mechanisms are distinct: the former arises from a topological mode in a locally resonant system, whereas the present mode emerges from geometry-dependent coupling in a dispersive linkage chain. The present mode is closely related to compact localized states in flat-band systems and other discrete lattices,[35, 36, 42, 43] as it likewise exhibits strictly finite spatial support in the linear regime. Unlike conventional flat-band compact localized states, however, the present state arises as the limiting case of a continuously tunable boundary edge mode in an otherwise dispersive chain. We validate the analytic extreme energy localization through both numerical simulations and experiments. The resulting mechanism provides a systematically controllable and experimentally accessible geometric route to compact boundary localization in mechanical metamaterials, with potential applications in impact mitigation, vibration isolation, and energy-guiding devices.

# 2 Results and Discussion

## 2.1 Geometry of the asymmetric linkage system

To create a linkage structure whose dynamics is primarily governed by geometry, we designed an asymmetric linkage unit cell as shown in Figure 1a. The unit cell consists of two rigid bars with unequal lengths, $2L$ and $2R$, connected by a revolute joint at the top hinge. Depending on which bar is longer, the unit cell is classified as either an $L$-type or an $R$-type configuration. A torsional spring with stiffness $k_\theta$ is placed at the top hinge, providing a restoring torque between the two bars. The right bottom hinge is fixed in translation but free to rotate, whereas the other bottom hinge translates along a horizontal rail, resulting in a single degree of freedom for the unit cell. The geometry of the unit cell is characterized by the angles $\theta^{\mathrm{L}}$ and $\theta^{\mathrm{R}}$, each defined as the angle between the bar and the reference vertical line. The clockwise direction is taken as positive for $\theta^{\mathrm{L}}$, whereas the counterclockwise direction is defined as positive for $\theta^{\mathrm{R}}$. Accordingly, in the configuration of Figure 1a, both $\theta^{\mathrm{L}}$ and $\theta^{\mathrm{R}}$ are positive. In contrast to linkage systems composed of two identical bars ($L = R$), the asymmetric linkage system admits two additional qualitatively distinct configuration states: *bulged-out* state and *nested-in* state. The bulged-out state corresponds to configurations in which both $\theta^{\mathrm{L}}$ and $\theta^{\mathrm{R}}$ are positive, while the nested-in state arises when $\theta^{\mathrm{L}}$ and $\theta^{\mathrm{R}}$ have opposite signs. For example, insets i and ii (also iii and iv) in Figure 1b have identical bar lengths but differ in the placement direction of the shorter bar, corresponding to the bulged-out and nested-in states, respectively. The existence of these two configuration states implies that $\theta$ of the shorter bar uniquely determines the configuration of the unit cell and can therefore be adopted as a generalized coordinate $q_{\mathrm{unit}} \in \mathbb{R}$ that provides a minimal and independent description of the system. Hence, the generalized coordinate of the unit cell is defined as

$$q_{\mathrm{unit}} = \begin{cases} \theta^{\mathrm{R}} & (L\text{-type}) \\ \theta^{\mathrm{L}} & (R\text{-type}) \end{cases} \tag{1}$$

where positive and negative values correspond to the bulged-out and nested-in states, respectively.

We define the length ratio of the two bars as $\Lambda = R/L$ and the rest angle between the bars as $\theta_0 = \theta^{\mathrm{L}} + \theta^{\mathrm{R}}$. Using these two parameters, $\Lambda$ and $\theta_0$, all possible configurations of the asymmetric linkage unit cell can be uniquely characterized. The configuration space shown in Figure 1b is divided into four distinct regions, indicated by different colors. First, the system can be classified into $L$-type or $R$-type configurations according to whether $\Lambda < 1$ or $\Lambda > 1$. Within each type, the configurations are further separated into bulged-out and nested-in states by the curve where the shorter bar is aligned with the vertical direction, i.e., where the generalized coordinate becomes zero. In each region of configuration space, a representative marker corresponding to a specific set of $\Lambda$ and $\theta_0$ values is shown, together with the associated linkage geometry. In total, four markers, denoted as *configurations i–iv*, are presented and will be used as representative cases throughout the following analysis. Notably, configurations i and iii, as well as configurations

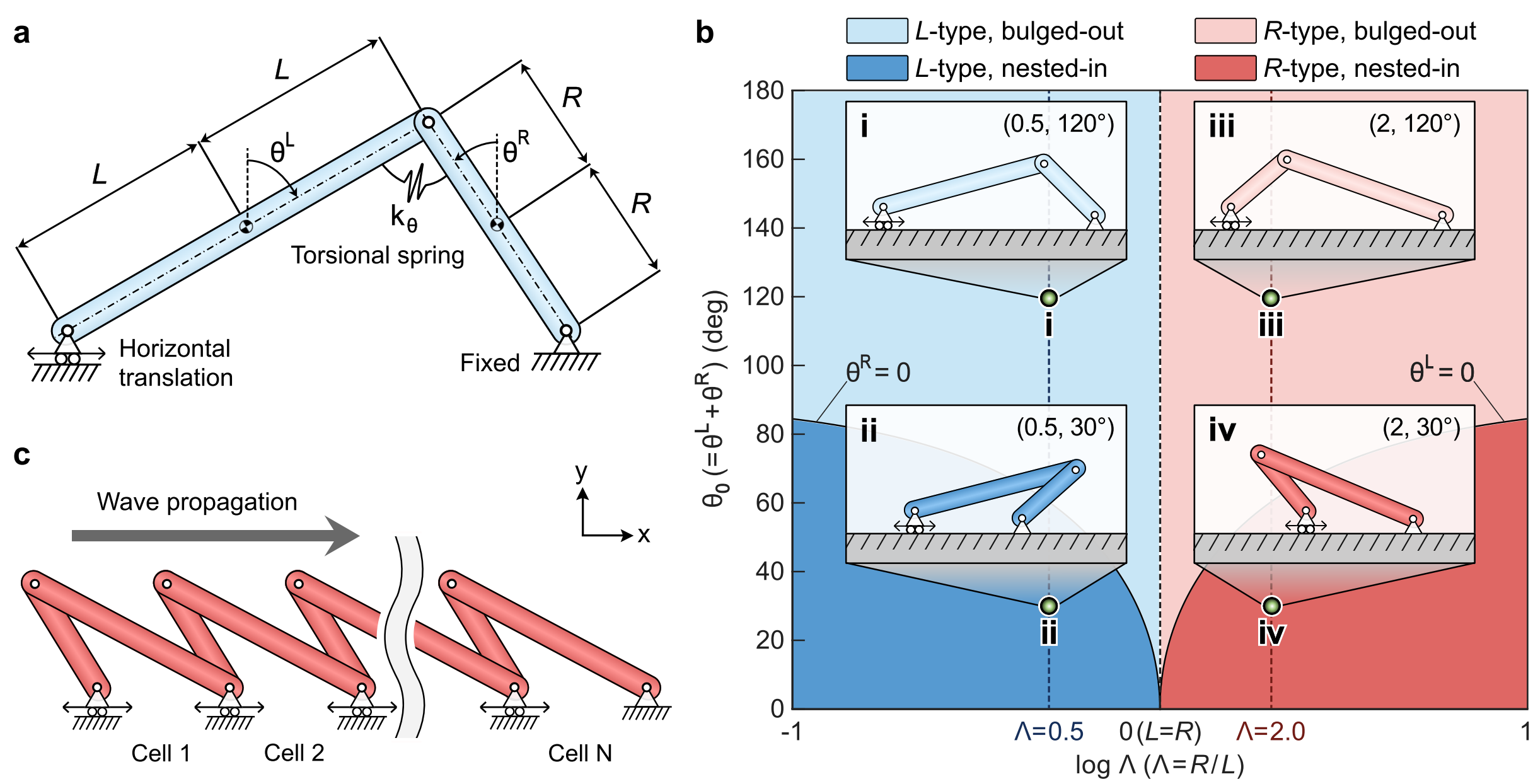

Figure 1: **Asymmetric linkage geometry defines four configuration regimes.** (a) Unit-cell geometry with a single degree of freedom. The torsional spring supports the equilibrium configuration at $\theta_0 = \theta_L + \theta_R$ and is omitted from subsequent schematics for clarity. (b) Configuration map in the $(\Lambda, \theta_0)$ parameter space, where $\Lambda = R/L$. The horizontal axis is plotted on a logarithmic scale to show the symmetry about $\log \Lambda = 0$. Markers i–iv identify representative configurations in the four geometric regimes, with the corresponding linkage geometries shown in the insets. (c) Chain of $N$ asymmetric linkage cells in the R-type, nested-in configuration. Mechanical waves propagate from the left free end toward the right fixed end along $+x$.

ii and iv, possess mirror-related unit-cell geometries and differ in their arrangement direction.

Using one of these types of unit cell, an asymmetric linkage chain can be constructed, as shown in Figure 1c by sequentially connecting unit cells from the rightmost fixed cell. This homogeneous chain is indexed using subscript $i$ from the left free end ($i = 1$) to the right fixed end ($i = N$), with the unit-cell coordinate denoted by $q_i$ and the full generalized-coordinate vector defined as $\vec{q} = [q_1, \ldots, q_N]^{\mathrm{T}} \in \mathbb{R}^N$. An initial excitation can be applied to the left free end, causing the first slider to move horizontally. This motion induces rotations of the connected linkages through the torsional spring and propagates sequentially through adjacent units. The fixed support at the right boundary restricts rigid-body translation of the chain, thereby allowing its intrinsic internal wave-propagation dynamics to be isolated and observed.

## 2.2 Configuration-dependent wave propagation

To describe the full dynamics of the present asymmetric linkage system, we first establish its governing equation of motion [24]:

$$\mathbf{G}^T\hat{\mathbf{M}}\mathbf{G}\ddot{\vec{q}} + \mathbf{G}^T\hat{\mathbf{M}}\dot{\mathbf{G}}\dot{\vec{q}} = \mathbf{G}^T\vec{f}. \tag{2}$$

Here, $\mathbf{G}(\vec{q}) \in \mathbb{R}^{6N\times N}$ is the configuration-dependent Jacobian matrix that maps the generalized-coordinate rates $\dot{\vec{q}} \in \mathbb{R}^{N\times 1}$ to the vector $\vec{v} \in \mathbb{R}^{6N\times 1}$ containing the translational and rotational velocities of the rigid bars, $\vec{v} = \mathbf{G}\dot{\vec{q}}$, while satisfying the free–fixed boundary condition. Because $\mathbf{G}$ depends nonlinearly on the system configuration, it introduces geometric nonlinearity into the governing equation. Accordingly, $\mathbf{G}^T\hat{\mathbf{M}}\mathbf{G} \in \mathbb{R}^{N\times N}$ represents a configuration-dependent effective mass matrix in the generalized-coordinate space, where $\hat{\mathbf{M}} \in \mathbb{R}^{6N\times 6N}$ contains the translational and rotational inertial properties of the individual rigid bars. The term $\mathbf{G}^T\hat{\mathbf{M}}\dot{\mathbf{G}}\dot{\vec{q}}$ is an additional velocity-dependent inertial term arising from the time variation of this configuration-dependent Jacobian. The vector $\mathbf{G}^T\vec{f} \in \mathbb{R}^{N\times 1}$ represents the generalized force associated with the elastic restoring forces from the torsional springs and the external forces. The detailed derivation of Equation (2) is provided in Supplementary Note 2.

Using this equation of motion, numerical simulations are conducted to investigate energy propagation within the system. We consider four different geometric configurations, denoted as configurations i–iv. The input conditions can be broadly classified into initial-displacement, initial-velocity, and external-force. Although all three input conditions exhibit similar qualitative trends, the main text focuses on the initial-velocity condition, which most clearly reveals the differences among the configurations. Simulation results under the other types of input conditions are provided in Supplementary Note 3. Specifically, an initial rotational velocity of $-18$ rad/s is imposed on the generalized coordinate of the first unit in a 20-cell system. Preliminary simulations confirmed that the consistent configuration-dependent trends persist at substantially smaller excitation amplitudes, for which the responses agree more closely with the linear modal predictions presented below. The present excitation amplitude was selected to demonstrate that these trends remain robust even in the presence of moderate nonlinear effects, while also providing a response magnitude comparable to that used in the experiments. The primary objective is to examine how distinct wave-propagation characteristics emerge depending on the geometric configuration.

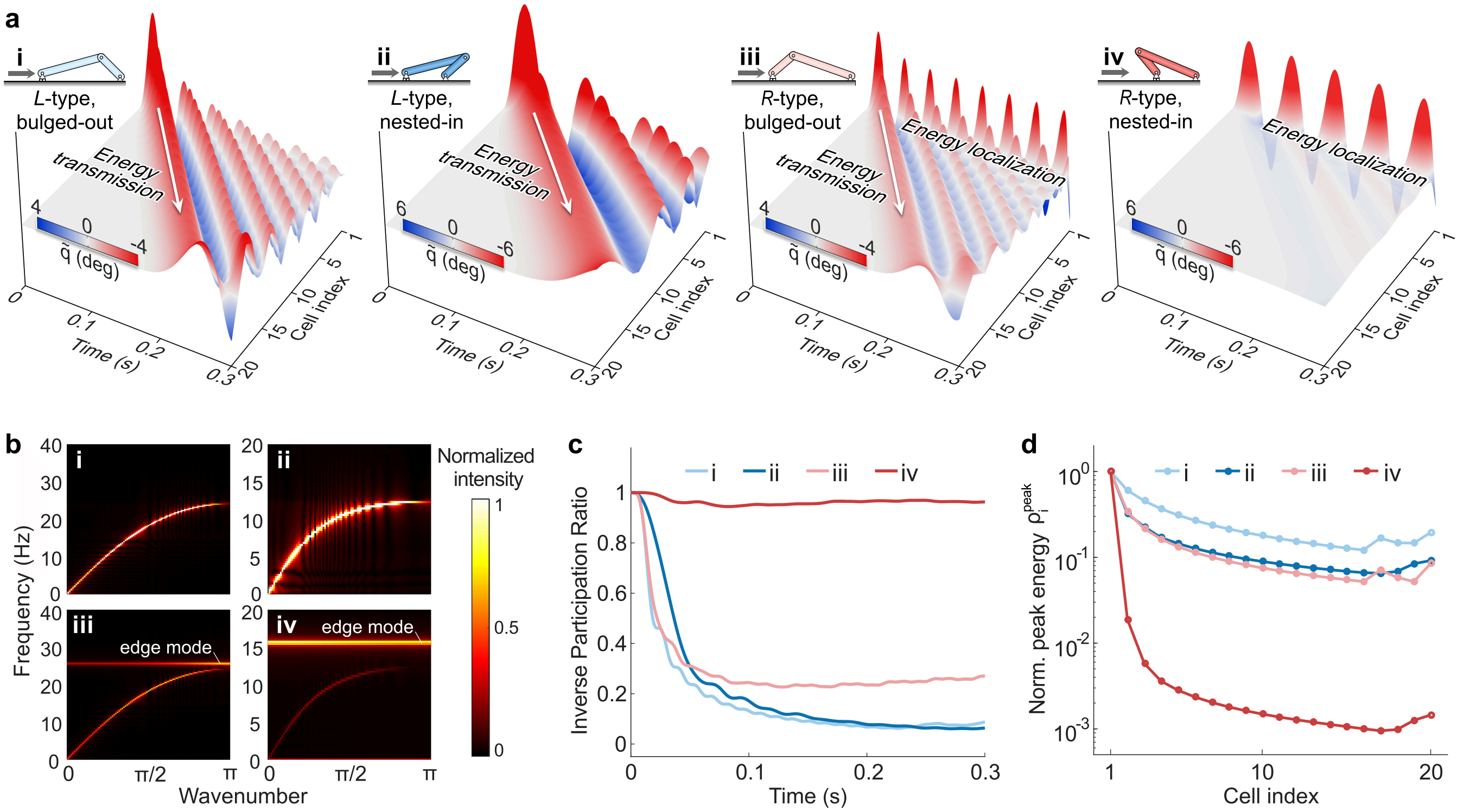

Figure 2: **Geometric configuration switches the system from bulk transmission to edge localization.** (a) Numerical time–space–displacement responses of the generalized-coordinate perturbation $\tilde{q}$ for configurations i–iv. The four geometries produce distinct degrees of wave transmission and boundary localization. (b) Corresponding two-dimensional Fourier spectra, calculated using simulations with spatial and temporal domains extended tenfold to improve spectral resolution. The nearly horizontal features in configurations iii and iv identify edge-mode responses. (c) Time evolution of the inverse participation ratio (IPR); larger values indicate stronger concentration of mechanical energy. (d) Cell-wise normalized peak energy on a semilogarithmic scale, showing the configuration-dependent spatial decay of transmitted energy.

In the numerical wave simulation results shown in Figure 2a, which depict the displacement of the generalized-coordinate perturbation $\tilde{q}$, each asymmetric linkage system exhibits qualitatively different dynamic responses. Although the $L$-type configurations, configurations i and ii, exhibit different propagation patterns, they consistently demonstrate clear bulk wave transmission. Configuration iii exhibits an intermediate response in which an edge-localized motion near the first cell coexists with appreciable bulk wave transmission along the chain. In contrast, configuration iv shows very weak wave propagation into the bulk, with the response dominated almost entirely by the edge mode.

We conduct a frequency-domain analysis to reveal the differences in wave propagation characteristics among the configurations. Accordingly, the frequency domain results shown in Figure 2b are obtained from a 200-cell system simulated for a duration ten times longer than that used in the corresponding time-domain simulations, improving both the wavenumber and frequency resolutions. Despite the increased system size, the edge modes in configurations iii and iv remain strongly localized, appearing as pronounced horizontal spectral lines extending over a broad range of wavenumbers. Indeed, this visualization suggests a connection between the relevant configuration and the broader concept of compact localized states [42], which has been explored in a variety of settings, including photonic, polaritonic, and electrical lattice chains [35, 43].

The effect of the presence of an edge mode can be interpreted as the spatial concentration of energy within the structure. To quantitatively characterize this energy localization, we introduce two scalar quantities. First, to evaluate the overall degree of localization, we use the inverse participation ratio (IPR), defined as

$$\mathrm{IPR}(t) = \frac{\sum_{i=1}^{N} E_i(t)^2}{\left(\sum_{i=1}^{N} E_i(t)\right)^2}, \tag{3}$$

where $E_i(t) = U_i(t) + T_i^{\mathrm{trans}}(t) + T_i^{\mathrm{rot}}(t)$ denotes the total mechanical energy associated with the $i$th unit cell, including the elastic potential energy stored in the torsional spring and the translational and rotational kinetic energies. The IPR ranges from $1/N$, corresponding to energy uniformly distributed over all $N$ cells, to 1, corresponding to complete localization within a single cell; thus, a higher IPR indicates stronger energy local-ization. In the time-varying IPR results shown in Figure 2c, the two $L$-type systems, configurations i and ii, show nearly identical IPR values, which are the lowest among all configurations, indicating minimal energy localization. In contrast, the $R$-type systems, configurations iii and iv, exhibit higher IPR values. In configuration iii, where an edge mode emerges alongside a modest degree of bulk propagation, the IPR attains values exceeding those of the $L$-type cases, indicating the onset of energy localization. Configuration iv, however, exhibits markedly higher IPR values compared with all other configurations.

Second, to identify where and to what extent energy localization occurs along the chain, we introduce the normalized peak energy per cell, defined as

$$\rho_i^{\text{peak}} = \frac{\max_{t\in[0,T]} E_i(t)}{\max_{t\in[0,T]} E_1(t)}. \tag{4}$$

This quantity provides a clearer measure of the spatial decay of energy by normalizing the maximum energy of each unit cell during the simulation by the peak energy of the first unit cell. As shown in Figure 2d, the normalized peak energy becomes lower in the order of configurations i, ii, iii, and iv. For configurations i, ii, and iii, the normalized peak energy at the distal end of the chain remains approximately on the order of $10^{-1}$, indicating that a finite amount of energy is still transmitted through the bulk. In contrast, configuration iv exhibits a much stronger spatial decay, with the normalized peak energy reduced to approximately the $10^{-3}$ level at the end of the chain.

These findings suggest the possibility of extreme energy localization in this simple linkage system. This observation naturally raises the question of how these differences in energy distribution arise and whether they can be systematically tuned. To address this question, an analytical modal approach will now be pursued.

## 2.3 Analytical framework of localization

The final equation of motion (Equation (2)) includes both the geometric relations of the asymmetric linkage unit cell and the cumulative structure under the free–fixed boundary condition. Consequently, direct application of conventional modal analysis, as in a standard mass–spring system, is not straightforward. To obtain a form suitable for modal analysis, the equation of motion is first linearized around the equilibrium configuration. Here, the equilibrium configuration is defined as the initial state specified by a point on the configuration map in Figure 1b, before any dynamic excitation is applied. This linearization yields

$$\bar{\mathcal{M}}\ddot{\tilde{\vec{q}}} + \bar{\mathbf{K}}\tilde{\vec{q}} = \vec{0}. \tag{5}$$

where $\tilde{\vec{q}}$ denotes the small displacement of the generalized-coordinate vector from equilibrium, $\vec{q}_0$. The linearized mass matrix is $\bar{\mathcal{M}} = \mathbf{G}^T\hat{\mathbf{M}}\mathbf{G}\big|_{\vec{q}=\vec{q}_0}$, while $\bar{\mathbf{K}}$ is obtained by linearizing $-\mathbf{G}^T\vec{f}$ about $\vec{q}_0$ and collecting the terms proportional to $\tilde{\vec{q}}$. The velocity-dependent term $\mathbf{G}^T\hat{\mathbf{M}}\dot{\mathbf{G}}\dot{\vec{q}}$ in Equation (2) does not contribute at first order because it is quadratic in the perturbation velocities.

Although this linearized form enables numerical eigenmode analysis, the resulting system matrices are fully populated due to the boundary constraints, making it difficult to derive closed-form modal solutions directly. We therefore introduce a coordinate transformation to express the system in a simplified form,

$$\vec{\phi} = \mathbf{L}\tilde{\vec{q}}, \quad \mathbf{L} = \begin{bmatrix} 1 & 1 & 1 & \cdots & 1 \\ 0 & 1 & 1 & \cdots & 1 \\ 0 & 0 & 1 & \cdots & 1 \\ \vdots & \vdots & \vdots & \ddots & \vdots \\ 0 & 0 & 0 & \cdots & 1 \end{bmatrix}. \tag{6}$$

The upper-triangular structure of $\mathbf{L}$ physically reflects the cumulative-coordinate transformation $\phi_i = \tilde{q}_i + \tilde{q}_{i+1} + \cdots + \tilde{q}_N$, where the transformed coordinate represents the configuration accumulated from the $i$-th cell to the last cell. The transformed eigenmodes can be chosen to be mutually orthogonal with respect to the transformed mass matrix introduced below. Substituting $\tilde{\vec{q}} = \mathbf{L}^{-1}\vec{\phi}$ into Equation (5) and premultiplying the resulting equation by $(\mathbf{L}^{-1})^T$ to preserve the symmetric matrix form leads to the transformed equation of motion,

$$\underbrace{(\mathbf{L}^{-1})^T\bar{\mathcal{M}}\mathbf{L}^{-1}}_{\tilde{\mathcal{M}}}\ddot{\vec{\phi}} + \underbrace{(\mathbf{L}^{-1})^T\bar{\mathbf{K}}\mathbf{L}^{-1}}_{\tilde{\mathbf{K}}}\vec{\phi} = \vec{0}. \tag{7}$$

The transformed mass and stiffness matrices, $\tilde{\mathcal{M}}$ and $\tilde{\mathbf{K}}$, exhibit a tridiagonal structure in the linearized regime, enabling a closed-form analysis of the modal characteristics.

In this transformed system, the tridiagonalization reduces the system matrices to their nearest-neighbor form, allowing a spatiotemporally separable modal representation whose distinct-frequency modes are orthogonal with respect to $\tilde{\mathcal{M}}$. Accordingly, the $n$th mode is expressed as

$$\phi_n(i) = a_n(i)e^{j\omega_n t}, \qquad a_n(i) = P_n r_n^{i-1} + Q_n s_n^{i-1}. \tag{8}$$

Here, $a_n(i)$ denotes the spatial amplitude of the $n$th mode at the $i$th transformed coordinate. The quantities $r_n$ and $s_n$ denote the two spatial factors of the $n$th mode associated with opposite spatial directions, while $P_n$ and $Q_n$ are their corresponding coefficients. Substituting this ansatz into Equation (7), together with the boundary equations, yields closed-form relations that determine the complete modal spectrum and the corresponding spatial mode shapes of the finite chain, including all bulk and edge modes (see Supplementary Note 4 for the detailed derivation). In particular, the bulk-mode frequency–wavenumber relation is

$$\omega_n^2 = \frac{4k_c \sin^2(\kappa_n/2)}{2\mathcal{G} - 2\gamma + \mathfrak{g} + 2(-\mathcal{G} + \gamma)\cos\kappa_n}, \tag{9}$$

where $\kappa_n$ denotes the wavenumber of the $n$th bulk mode. The edge-mode frequency and spatial decay ratio are, respectively,

$$\omega_{\text{e}}^2 = \frac{k_c\mathfrak{g}}{\mathcal{G}\mathfrak{g} - \gamma^2}, \tag{10}$$

$$r_{\text{e}} = \frac{\gamma}{\gamma - \mathfrak{g}}. \tag{11}$$

The subscript e denotes quantities associated with the edge mode. The coefficients appearing in Equations (9)–(11) are type-dependent and can be expressed in terms of the geometric and inertial parameters introduced in Figure 1, as summarized in Table 1. Here, $m^{\mathrm{L}}$ and $m^{\mathrm{R}}$ denote the masses of the left and right bars, respectively, whereas $J^{\mathrm{L}}$ and $J^{\mathrm{R}}$ denote their corresponding mass moments of inertia about the bar centers.

For a semi-infinite chain, boundedness requires $Q_{\mathrm{e}} = 0$, such that the transformed edge-mode profile is governed solely by the decaying spatial factor $r_{\mathrm{e}}$. This localization is preserved in the original generalized coordinate: because $\widetilde{q}_i = \phi_i - \phi_{i+1}$, the profile $\phi_i = P_{\mathrm{e}} r_{\mathrm{e}}^{i-1}$ yields $\widetilde{q}_i = P_{\mathrm{e}}(1 - r_{\mathrm{e}}) r_{\mathrm{e}}^{i-1}$. Thus, the transformed and original coordinates exhibit the same spatial decay ratio, confirming that the edge-mode solution is localized in both coordinate systems.

For a finite chain, Equations (10) and (11) accurately describe the edge mode when its localization length is much shorter than the chain length. In this regime, the influence of the distant boundary is exponentially small, and the coefficient $Q_{\mathrm{e}}$ of the growing spatial component is effectively zero (see Supplementary Note 4).

Meanwhile, the spatial attenuation of the edge mode can be quantified by extending the bulk-mode dispersion relation in Equation (9) to the evanescent regime, where the complex wavenumber is decomposed as $\kappa_n = \kappa_{n,\mathrm{real}} + j\kappa_{n,\mathrm{imag}}$ into its real and imaginary parts. Omitting the mode index $n$, Equation (9) can be expressed in terms of $\kappa_{\mathrm{imag}}$, yielding the energy localization length $L_{\mathrm{energy}}$ as follows:

$$\kappa_{\mathrm{imag}} = \operatorname{arccosh}\left|\frac{2k_c - (2\mathcal{G} - 2\gamma + \mathfrak{g})\omega^2}{2k_c + 2(-\mathcal{G} + \gamma)\omega^2}\right|, \quad L_{energy} = \frac{1}{2\kappa_{\mathrm{imag}}} \tag{12}$$

If a nonsingular edge mode exists in a given system, its frequency intersects the evanescent branch at a finite value of $\kappa_{\mathrm{imag}}$. A larger value of $\kappa_{\mathrm{imag}}$ at this intersection corresponds to a shorter localization length and, consequently, stronger edge localization. This correspondence shows that the evanescent extension of the bulk dispersion directly captures edge localization.

Based on this analysis, the overall modal characteristics of configurations i–iv can be examined. Figure 3a presents the corresponding modal frequencies for each configuration. The simulation-based frequencies, shown in black, are extracted from the Fourier spectrum results in Figure 2b. In contrast, the blue and red markers represent the analytical bulk- and edge-mode frequencies, respectively, obtained from the closed-form expressions in Equations (9) and (10). These analytical results also exactly coincide with the numerical eigenfrequencies obtained from the linearized equation of motion (Equation (5)) using the corresponding system parameters. The agreement among the simulation results, numerical eigenmode analysis, and analytical approach confirms the validity of the derived solutions. It is worth noting that the modal analysis is performed under the assumption of linearization, whereas the simulations may include nonlinear effects. Nevertheless, within the considered excitation range, the simulation results agree closely with the analytical predictions, with relative differences of less than 1%. This close agreement indicates that the predicted modal characteristics remain robust even when the system response extends slightly beyond the linear regime.

Figure 3b compares the selected mode shapes of configurations i–iv, demonstrating that the modal behavior strongly depends on the system configuration. For a consistent comparison with the analytical solutions, all numerical eigenvectors and simulated displacement profiles are transformed into the $\boldsymbol{\phi}$ coordinate. The analytical mode shapes, represented by the blue and red markers, are obtained from the closed-form ansatz in Equation (8), and agree with the transformed numerical eigenvectors obtained from the linearized equation of motion in Equation (5). The black dashed curves represent the corresponding transformed simulation results obtained by applying a sinusoidal excitation to the first cell at the relevant modal frequency and plotting the normalized displacement profile after a stable standing-wave response has developed (see Supplementary Note 4 for details of the procedure). Configurations i and ii support only bulk modes with sinusoidal profiles throughout the modal spectrum. The fifth mode of configuration i and the tenth mode of configuration ii both show good agreement between simulation and analysis, with slightly closer agreement for the lower-frequency fifth mode. The small discrepancy at the higher frequency arises from the flatter dispersion and closer modal spacing, which enhance neighboring-mode contributions. In contrast, the last modes of configurations iii and iv are edge modes exhibiting staggered (negative $r$) and non-staggered (positive $r$) exponential decay, respectively. For both configurations, the mode shapes obtained under sinusoidal excitation agree closely with the analytical and numerical eigenmode predictions. For further comparison, the RMS-normalized displacement profiles obtained from the initial-velocity simulations in Figure 2a, likewise transformed into the $\boldsymbol{\phi}$ coordinate, are also overlaid on the panels of iii and iv. Unlike the sinusoidal excitation, which selectively probes a prescribed modal frequency, the localized initial-velocity input primarily reflects the spatial modal content coupled to the excitation site. Accordingly, the resulting RMS profile depends strongly on the degree to which the relevant edge mode is concentrated near the first cell. For configuration iii, the relatively large magnitude of the decay factor, $|r| = 0.4483$, produces a spatial profile that extends over several cells. The RMS displacement at the second cell is approximately 0.8462, substantially larger than the corresponding analytical edge-mode amplitude of 0.4483, while non-negligible bulk responses remain throughout the chain. This indicates that, even under a localized input, a system with a relatively large $|r|$ can significantly excite bulk modes alongside the edge mode. In contrast, the smaller magnitude of the decay factor for configuration iv, $|r| = 0.0669$, produces a strongly localized spatial profile. Although the RMS displacement at the second cell, approximately 0.1488, is somewhat larger than the analytical value of 0.0669, the overall response remains strongly confined near the excited boundary, with an RMS displacement of only approximately 0.0152 at the twentieth cell. Thus, the strong spatial concentration of the edge mode near the excitation site allows it to dominate the response, resulting in closer agreement with the analytical edge-mode profile and more effective energy localization.

Table 1: Type-dependent coefficients of the transformed equation of motion

| **Coeff.** | ***L*-type** | ***R*-type** |
|---|---|---|
| $k_c$ | $k_\theta\left(1+\dfrac{R\sin\theta^{\mathrm{R}}}{L\sin\theta^{\mathrm{L}}}\right)^2$ | $k_\theta\left(1+\dfrac{L\sin\theta^{\mathrm{L}}}{R\sin\theta^{\mathrm{R}}}\right)^2$ |
| $\mathcal{G}$ | $R^2m^{\mathrm{L}}\left[\left(\dfrac{\sin\theta^{\mathrm{R}}}{\tan\theta^{\mathrm{L}}}+2\cos\theta^{\mathrm{R}}\right)^2+\sin^2\theta^{\mathrm{R}}\right]$ $+J^{\mathrm{L}}\left(\dfrac{R\sin\theta^{\mathrm{R}}}{L\sin\theta^{\mathrm{L}}}\right)^2+R^2m^{\mathrm{R}}+J^{\mathrm{R}}$ | $L^2m^{\mathrm{L}}\left[\left(2\dfrac{\sin\theta^{\mathrm{L}}}{\tan\theta^{\mathrm{R}}}+\cos\theta^{\mathrm{L}}\right)^2+\sin^2\theta^{\mathrm{L}}\right]$ $+J^{\mathrm{L}}+L^2m^{\mathrm{R}}\left(\dfrac{\sin^2\theta^{\mathrm{L}}}{\sin^2\theta^{\mathrm{R}}}\right)+J^{\mathrm{R}}\left(\dfrac{L\sin\theta^{\mathrm{L}}}{R\sin\theta^{\mathrm{R}}}\right)^2$ |
| $\gamma$ | $2R^2\left(\dfrac{\sin\theta^{\mathrm{R}}}{\tan\theta^{\mathrm{L}}}+\cos\theta^{\mathrm{R}}\right)$ $\times\left[m^{\mathrm{L}}\left(\dfrac{\sin\theta^{\mathrm{R}}}{\tan\theta^{\mathrm{L}}}+2\cos\theta^{\mathrm{R}}\right)+m^{\mathrm{R}}\cos\theta^{\mathrm{R}}\right]$ | $2L^2\left(\dfrac{\sin\theta^{\mathrm{L}}}{\tan\theta^{\mathrm{R}}}+\cos\theta^{\mathrm{L}}\right)$ $\times\left[m^{\mathrm{L}}\left(2\dfrac{\sin\theta^{\mathrm{L}}}{\tan\theta^{\mathrm{R}}}+\cos\theta^{\mathrm{L}}\right)+m^{\mathrm{R}}\left(\dfrac{\sin\theta^{\mathrm{L}}}{\tan\theta^{\mathrm{R}}}\right)\right]$ |
| $\mathfrak{g}$ | $4R^2\left(\dfrac{\sin\theta^{\mathrm{R}}}{\tan\theta^{\mathrm{L}}}+\cos\theta^{\mathrm{R}}\right)^2(m^{\mathrm{L}}+m^{\mathrm{R}})$ | $4L^2\left(\dfrac{\sin\theta^{\mathrm{L}}}{\tan\theta^{\mathrm{R}}}+\cos\theta^{\mathrm{L}}\right)^2(m^{\mathrm{L}}+m^{\mathrm{R}})$ |

Figure 3c quantifies the spatial attenuation associated with the edge modes by plotting the evanescent branch of $\kappa_{\mathrm{imag}}$. For configurations i and ii, the evanescent branch emerges above the highest bulk-mode frequency. However, because these configurations possess no eigenfrequency intersecting this branch, they do not support a physical mode with spatially decaying amplitude. In contrast, configurations iii and iv exhibit an edge-mode frequency that intersects the evanescent branch. At each intersection, $\kappa_{\mathrm{imag}}$ and the corresponding energy localization length $L_{\mathrm{energy}}$ defined in Equation (12) are uniquely determined, as shown in Figure 3c. The substantially shorter localization length of configuration iv is consistent with its more strongly confined edge-mode profile (Figure 3b) and with the energy characteristics observed in the simulations (Figure 2c, d). These results provide a quantitative description of the spatial localization of the edge mode and its connection to the bulk spectral characteristics.

#### 2.3.1 Condition for edge mode existence

As demonstrated by the simulations and modal analysis, edge modes emerge only in configurations iii and iv. This naturally raises the question of why such modes occur exclusively in the *R*-type systems. A localized edge-mode solution requires $|r| < 1$. Accordingly, the expression for $r$ in Equation (11) must satisfy the following condition:

$$-1 < \frac{\gamma}{\gamma - \mathfrak{g}} < 1. \tag{13}$$

Supplementary Note 5 shows that this condition is satisfied for the *R*-type configuration, whereas it is never satisfied for the *L*-type configuration. Therefore, while the *L*-type configuration cannot physically support an edge mode, every physically admissible *R*-type configuration possesses such a mode.

### 2.4 Singular edge mode

The edge-mode spatial decay ratio $r_{\mathrm{e}}$ is governed by the geometric and inertial parameters of the system through Equation (11). Under the uniform-density assumption, the contour lines in the $(\Lambda, \theta_0)$ configuration space shown in Figure 4a represent continuous sets of configurations having the same value of $r_{\mathrm{e}}$. In particular, the red contour line in the *R*-type, nested-in region defines the locus satisfying $r_{\mathrm{e}} = 0$. In this limiting state, only the first unit cell has a nonzero amplitude, whereas all subsequent cells have strictly zero amplitudes in the linear regime. We refer to this compactly supported boundary state as the singular edge mode, following terminology in topological metamaterials [34]. The singular edge mode shares the defining compact-support property of compact localized states, while arising through a distinct boundary-localization mechanism. Its singular nature is further clarified below through the divergence of the imaginary wavenumber.

From Equation (11), the singular condition $r_{\mathrm{e}} = 0$ requires $\gamma = 0$ with $\mathfrak{g} \neq 0$. Substituting the *R*-type expression for $\gamma$ yields an explicit closed-form condition for the existence of the singular edge mode (see Supplementary Note 6 for the detailed derivation). For the uniform-density bars considered here, the mass ratio satisfies $m_{\mathrm{R}}/m_{\mathrm{L}} = R/L = \Lambda$, and the condition becomes

$$\frac{\tan\theta^{\mathrm{R}}}{\tan\theta^{\mathrm{L}}} = -(\Lambda + 2). \tag{14}$$

Equation (14) provides a direct closed-form design criterion for realizing the singular edge mode solely from the unit-cell parameters, without requiring a numerical search over the configuration space.

The singular edge mode can also be interpreted in terms of the spatial attenuation characterized by $\kappa_{\mathrm{imag}}$. According to Equation (12), the evanescent branch develops a vertical asymptote when its denominator vanishes, yielding

$$\omega_\infty^2 = \frac{k_c}{\mathcal{G} - \gamma}, \tag{15}$$

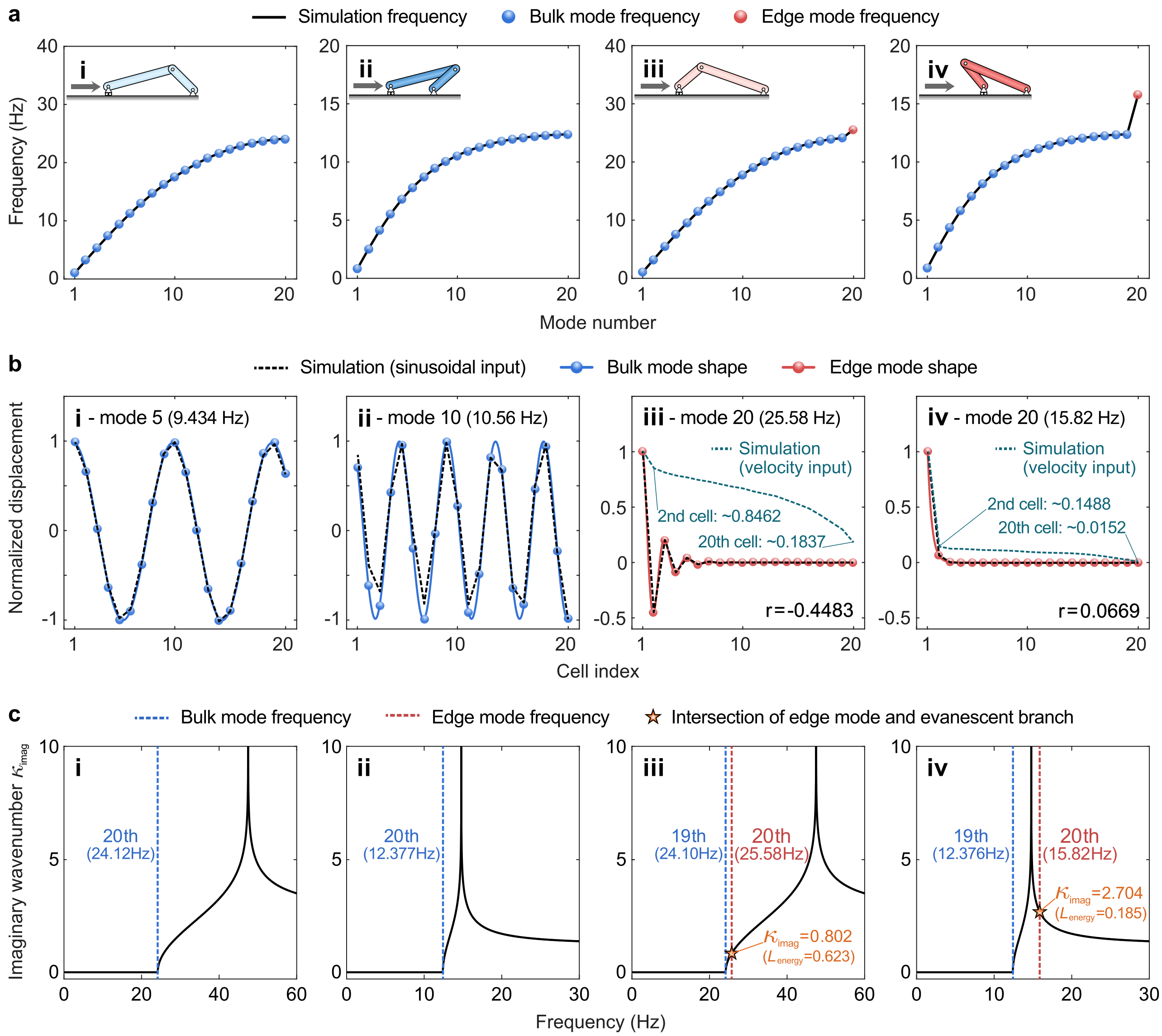

Figure 3: **Closed-form modal analysis explains configuration-dependent edge localization.** (a) Modal frequencies for configurations i–iv. Black curves show frequencies extracted from the simulated Fourier spectra in Figure 2b, whereas blue and red markers denote the analytical bulk- and edge-mode frequencies, respectively. (b) Selected mode shapes for configurations i–iv in the transformed $\phi$ coordinate. Black curves show simulated responses under sinusoidal excitation at the corresponding modal frequencies, and blue and red markers show the analytical bulk- and edge-mode shapes. For configurations iii and iv, RMS-normalized profiles obtained from the initial-velocity simulations (Figure 2a) are also shown in the $\phi$ coordinate. (c) Evanescent branches described by the imaginary wavenumber $\kappa_{\text{imag}}$. Blue dashed lines mark the highest bulk-mode frequencies, and red dashed lines mark edge-mode frequencies where present. For configurations iii and iv, the intersections with the evanescent branches determine the energy localization length.

where $\omega_\infty$ denotes the frequency of infinite spatial attenuation. The presence of this asymptote alone, however, does not ensure that it corresponds to a physical eigenmode of the finite chain. Complete suppression of spatial penetration requires the edge-mode frequency to coincide with $\omega_\infty$. Using Equation (10), this condition gives

$$\omega_e^2 = \omega_\infty^2 \iff \frac{k_c \mathfrak{g}}{\mathcal{G}\mathfrak{g} - \gamma^2} = \frac{k_c}{\mathcal{G} - \gamma} \iff \gamma(\gamma - \mathfrak{g}) = 0. \quad (16)$$

The root $\gamma = \mathfrak{g}$ makes the decay ratio $r_e = \gamma/(\gamma - \mathfrak{g})$ undefined, leaving $\gamma = 0$ as the physically admissible solution and hence $r_e = 0$. Thus, the imaginary-wavenumber interpretation independently recovers the same singular condition given by Equation (14).

For a rest angle of $\theta_0 = 30°$, the singular condition is satisfied at $\Lambda \approx 1.421$, which defines configuration v. Its corresponding geometry is illustrated in the inset of Figure 4a. For configuration v, we perform a numerical simulation (Figure 4b) using the same excitation condition as in the preceding simulations. Results under other excitation conditions are provided in Supplementary Note 7.

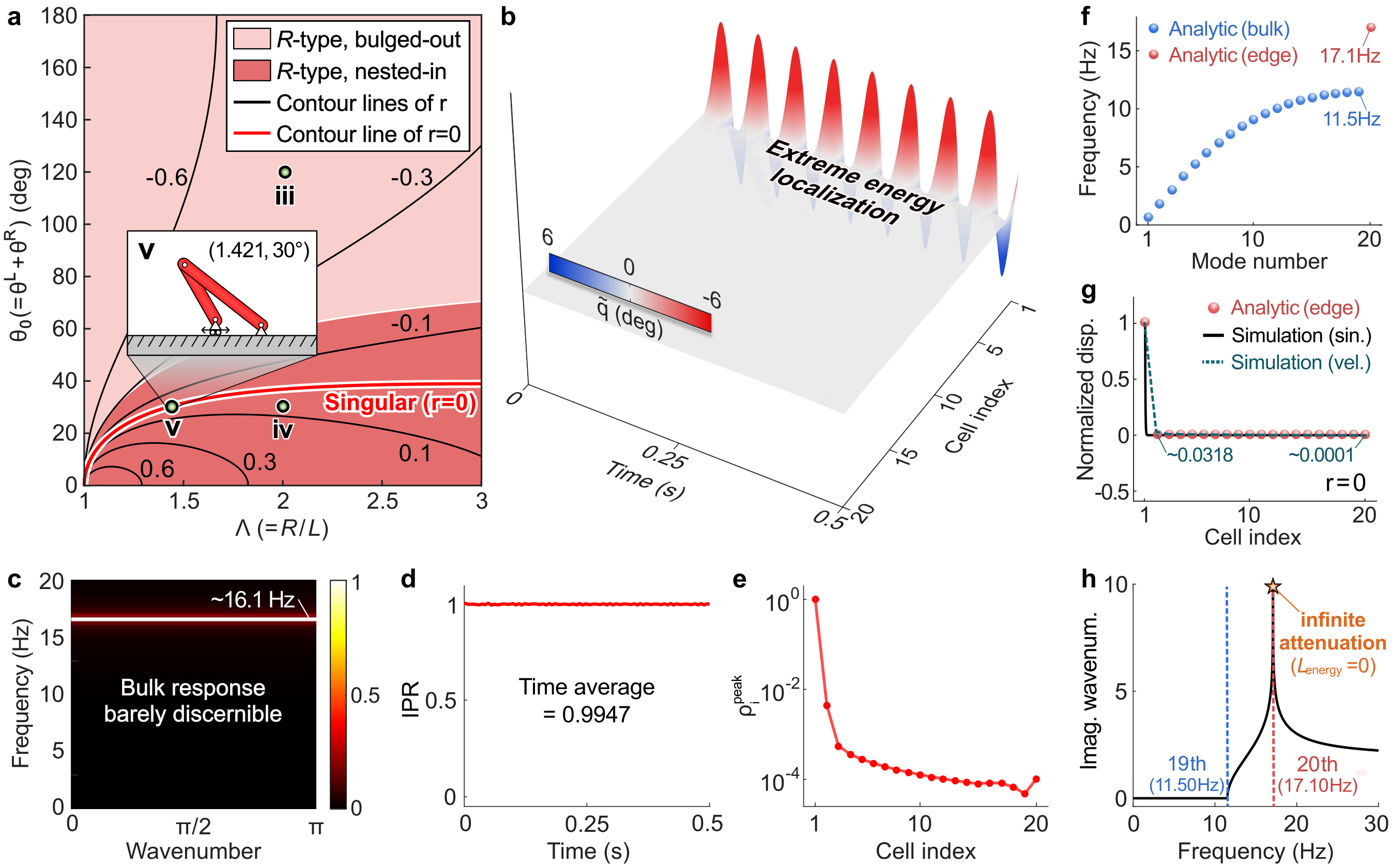

Figure 4: **A singular edge mode enables extreme localization at the boundary.** (a) Contour map of the edge-mode spatial decay ratio $r_e$ in the $(\Lambda, \theta_0)$ configuration space. The highlighted singular line satisfies $r_e = 0$; configuration v lies on this line, and its geometry is shown in the inset. (b) Numerical wave response of configuration v, showing extreme localization near the free boundary. The simulation is extended to 0.5 s so that the weak bulk response can reach the distal end. (c) Fourier spectrum showing a pronounced, nearly horizontal edge-mode feature and a barely discernible bulk response. (d) IPR as a function of time, approaching the single-cell localization limit. (e) Cell-wise normalized peak energy on a semilogarithmic scale. (f) Modal frequencies of configuration v. (g) Analytical singular edge-mode shape together with normalized profiles in the $\phi$ coordinate obtained under sinusoidal and initial-velocity excitation. (h) Evanescent branch of $\kappa_{\text{imag}}$. The edge-mode frequency coincides with the vertical asymptote, for which $\kappa_{\text{imag}} \to \infty$ and $L_{\text{energy}} \to 0$, corresponding to formally infinite spatial attenuation in the linear model.

The simulation shows that the response of cells beyond the first cell is strongly suppressed. This time-domain localization is consistently supported by the spectral and energy-based analyses in Figure 4c–e. In Figure 4c, the edge-mode contribution appears as a pronounced, nearly horizontal spectral feature, whereas the bulk response is difficult to discern. The IPR approaches unity, with a time-averaged value of 0.9947 (Figure 4d), and the normalized peak energy decreases to approximately $10^{-4}$ at the end of the chain (Figure 4e). Together, these results provide direct quantitative evidence that the singular edge mode induces extreme energy localization.

The closed-form modal relations yield the modal spectrum and mode shape of configuration v, as shown in Figure 4f, g. The modal spectrum contains one edge mode, as in the other $R$-type systems. The edge-mode frequency is approximately 17.1 Hz, whereas the dominant frequency of the finite-amplitude simulation, represented by the nearly horizontal feature in Figure 4c, is approximately 16.1 Hz. This corresponds to a relative difference of about 5.8%. As shown in Supplementary Note 7, the modal-analysis frequency and the dominant simulation frequency agree closely in the low-amplitude regime, whereas a downward frequency shift emerges as the excitation amplitude increases. This behavior is consistent with amplitude-dependent nonlinear effects arising from the configuration-dependent inertia and nonlinear restoring response.

The negligible bulk-mode contribution observed in the simulations can be understood from the spatial structure of the singular edge mode. As shown in Figure 4g, the analytical mode shape has a nonzero amplitude only at the first cell, whereas all subsequent cells have exactly zero modal amplitudes. The spatial profile obtained under sinusoidal excitation at the singular edge-mode frequency agrees closely with the analytical prediction. For comparison, the RMS-normalized profile obtained from the initial-velocity simulation in Figure 4b is also shown as the dashed curve. Although this excitation is not frequency-selective, the normalized amplitude is only approximately 0.0318 at the second cell and approximately 0.0001 at the last cell. Thus, under both excitation conditions, the response remains almost entirely confined to the first cell. The boundary-localized excitation couples strongly to the singular edge mode, allowing it to dominate the response and substantially suppress bulk-mode participation. This explains why the dynamic response of configuration v is governed almost exclusively by the singular edge mode, despite the existence of bulk modes in the theoretical modal structure of the system.

Figure 4h shows the evanescent branch for configuration v. The edge-mode frequency coincides with its vertical asymptote, such that $\kappa_{\text{imag}}$ formally diverges and $L_{\text{energy}}$ approaches zero. Unlike configurations iii and iv, whose edge-mode frequencies intersect the evanescent branches at finite values of $\kappa_{\text{imag}}$ and therefore retain exponentially decaying tails, configuration v intersects the branch precisely at the infinite attenuation frequency. Therefore, the ideal singular edge mode has no decaying tail and is confined entirely to the first cell.

## 2.5 Experimental observations

To experimentally examine the configuration-dependent wave propagation behavior and energy localization effects analyzed previously for configurations i–v, we conducted dynamic experiments. The experimental setup shown in Figure 5a was designed to realize the desired linkage geometry and constraints. Each unit cell was 3D-printed using PLA filament and mounted on a sliding base equipped with linear bushings, allowing it to slide along parallel shafts. Markers attached to the sliding bases were tracked using a camera to capture the spatiotemporal wave propagation behavior. Three types of unit cells were fabricated according to their geometric configurations, and torsional springs were embedded inside each unit cell to provide the restoring stiffness. Excitation was applied by imposing an initial angular displacement of 30° on the first cell and then releasing it. The key experimental specifications are summarized in the Methods, while a detailed description is provided in Supplementary Note 8.

The experimentally observed responses shown in Figure 5b exhibit trends consistent with the simulation results presented in Figure 2a and 4b for each configuration. Configurations i and ii show clear energy transmission along the chain, whereas configurations iv and v exhibit pronounced energy localization, with configuration v showing near-complete localization. These configuration-dependent trends are also reflected in the experimental Fourier spectra provided in Supplementary Note 8.

For the bulged-out mirror-related pair, configurations i and iii, the difference in the experimental responses is relatively subtle. Configuration iii shows a slightly stronger tendency for the motion to remain near the excitation side than configuration i; however, within the resolution of the present experiment, the edge-mode contribution cannot be clearly distinguished from the coexisting bulk response. By contrast, the difference is much more pronounced for the nested-in mirror-related pair, configurations ii and iv. In configuration ii, wave propagation remains dominant over multiple unit cells, whereas configuration iv shows significant displacement only within the first few cells, indicating strong localization near the excitation site. Most notably, configuration v exhibits extreme energy localization, with appreciable displacement occurring predominantly in the first cell.

These qualitative trends are further supported by the quantitative localization measures shown in Figure 5c, d. The IPR results in Figure 5c show that configurations i and ii maintain relatively low IPR values, consistent with energy transmission over multiple cells. In contrast, configurations iv and v exhibit substantially higher IPR values, indicating that the experimentally measured response remains strongly localized near the excitation site. The fluctuations away from unity occur primarily during intervals in which the measured first-cell energy becomes small, making the calculated IPR particularly sensitive to tracking noise, finite resolution, and measurement errors amplified by experimental damping and wave attenuation. The cell-wise normalized peak energy in Figure 5d further supports the interpretation. Configurations i and ii retain relatively high normalized peak energy near the distal end of the chain, remaining above the $10^{-2}$ level, while the normalized peak energy decreases more strongly in configurations iii, iv, and v as the cell index increases. In particular, the normalized peak energy falls below the $10^{-4}$ level in configuration iv and reaches approximately the $10^{-5}$ level in configuration v. Together, these measurements reproduce the overall propagation and localization trends observed in the simulations despite the dissipation and finite-amplitude effects discussed in Supplementary Note 8, demonstrating that the configuration-dependent localization remains robust under nonideal physical conditions.

# 3 Conclusion

In this study, we developed an asymmetric linkage system and investigated its energy-transfer and localization characteristics through numerical simulations, modal analysis, and experiments. The system is classified into four representative configurations according to the unit-cell geometry. Numerical simulations performed on these configurations show that the dynamic behavior of the chain strongly depends on the prescribed configuration. In particular, by comparing mirror-related configuration pairs with identical bar lengths and rest angles but opposite arrangement directions, we show that reversing the orientation of the chain relative to the boundary can switch the system between bulk transmission and edge-mode-induced localization. Specifically, L-type chains favor bulk transmission from the left free boundary, whereas R-type chains support an edge mode and therefore exhibit stronger boundary localization. Within the R-type class, the nested-in configuration is more favorable for localization and produces substantially stronger confinement than the bulged-out configuration.

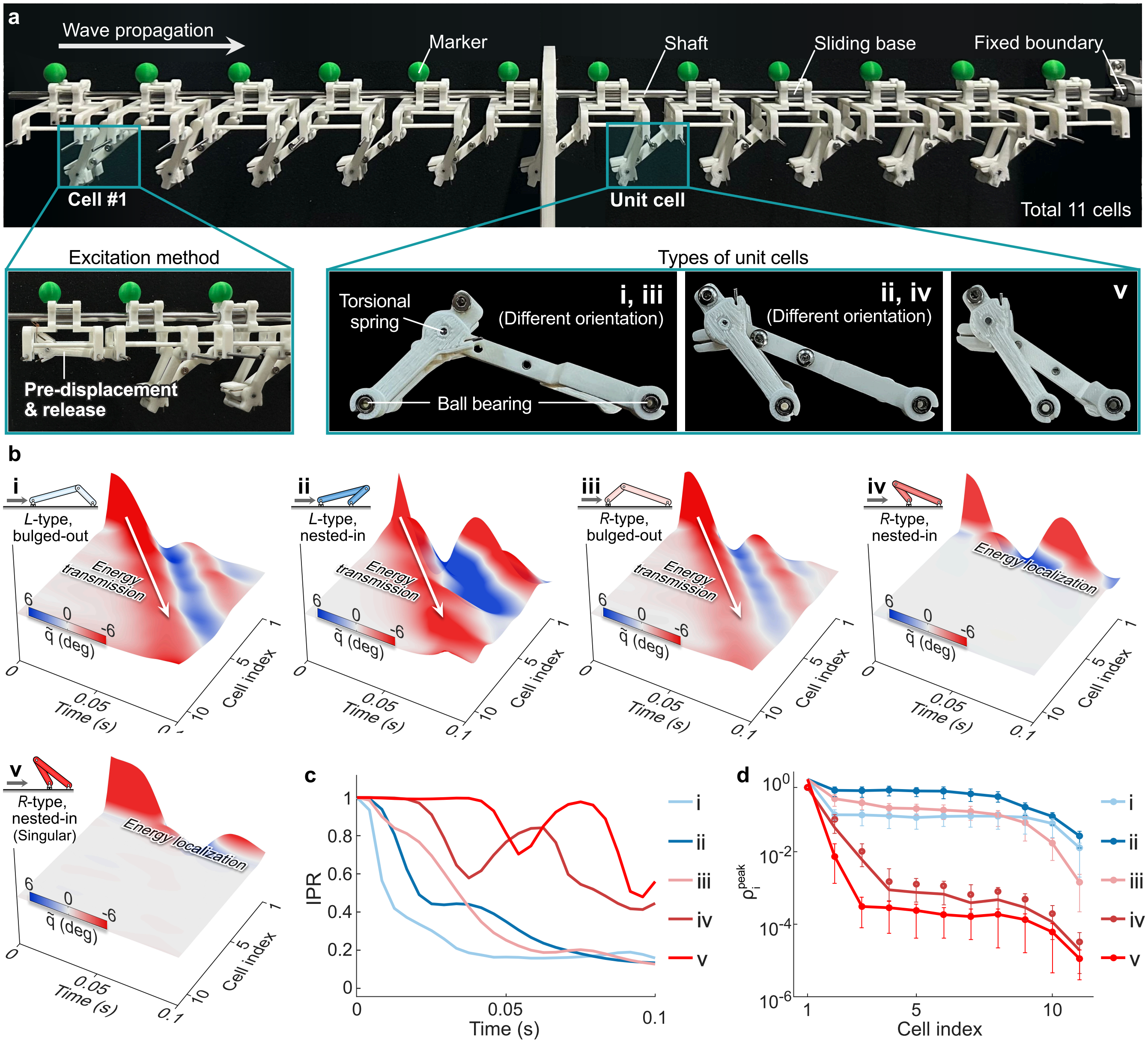

Figure 5: **Experiments reproduce the configuration-dependent wave propagation.** (a) Experimental apparatus and interchangeable linkage components used to realize configurations i–v. The response is initiated by imposing an initial displacement on the first cell and releasing it. (b) Measured time–space–displacement responses of the generalized coordinate for configurations i–v. (c) Time evolution of the IPR averaged over five experimental trials. (d) Cell-wise normalized peak energy on a semilogarithmic scale. Markers denote the mean over five trials, and error bars indicate the full range.

To uncover the underlying principle behind this behavior, we developed a modal analysis framework based on the transformation of the system matrices through linearization and tridiagonalization. This formulation enables a closed-form description of the mode structure, allowing the edge mode to be interpreted in terms of spatial decay. By extending the bulk-mode dispersion relation into the evanescent regime, the imaginary part of the wavenumber provides a quantitative connection between the bulk spectral characteristics and edge localization. In particular, the intersection between the edge-mode frequency and the evanescent branch determines the spatial attenuation rate and the corresponding energy localization length. Therefore, the proposed analysis is not limited to the present asymmetric linkage system and may also provide a useful framework for understanding localized modes in broader classes of discrete mechanical systems.

This analysis reveals that a specific nested-in configuration of the present system supports a singular edge mode, which acts as a mechanism for extreme energy localization. In the ideal analytical mode, only the first unit cell exhibits a nonzero amplitude, while all subsequent cells have exactly zero amplitude, resulting in complete spatial confinement. From the imaginary-wavenumber perspective, this state represents the limiting case in which the edge-mode frequency coincides with the singular attenuation frequency, such that $\kappa_{\text{imag}} \to \infty$ and $L_{\text{energy}} \to 0$. Based on the modal analysis, we analytically derived the condition for the emergence of this singular edge mode. Beyond the condition itself, the closed-form expressions for the modal frequency and spatial decay show that both the operating frequency and the degree of localization can be tuned through the unit-cell geometry. Finally, to examine the predicted configuration-dependent energy transfer and localization experimentally, we constructed an experimental setup and tested five configurations, comprising one representative from each of the four regions of configuration space and the singular state, demonstrating distinct localization and propagation trends, in line with our theoretical predictions.

Future work will involve extending the proposed linkage system in several directions, particularly by incorporating nonlinear aspects of its dynamic response. This effort will include analyzing the nonlinear wave solutions and examining the associated breather states [44, 45] and their stability [46], investigating amplitude-dependent nonlinear continuation of the localized modes, and exploring the effects of various boundary conditions on energy transfer and localization. Another important direction, motivated by the experimental results, is to quantify the role of dissipation in the linkage model, so as to obtain a closer correspondence between experimental findings and numerical simulations. Furthermore, introducing additional intercell springs at the bottom hinges provides a route to topologically nontrivial configurations, while spatially nonuniform unit-cell arrangements could serve as another design direction for controlling energy transfer and localization. Overall, the asymmetric linkage system proposed in this study provides a minimal and analytically tractable platform for realizing extreme energy localization, with potential applications in impact mitigation, vibration isolation, and programmable energy-guiding mechanical metamaterials.

# 4 Methods

## 4.1 Numerical simulations

Numerical simulations were performed in MATLAB by integrating Equation (2), using the adaptive Runge–Kutta solver `ode45`. For all simulations, the combined bar length, $2L + 2R$, and total bar mass were fixed at 150 mm and 30 g, respectively, with a uniform linear mass density. The corresponding mass moments of inertia about the bar centers were calculated as $J^{\text{L}} = \frac{1}{3} m^{\text{L}} L^2$ and $J^{\text{R}} = \frac{1}{3} m^{\text{R}} R^2$. The torsional stiffness was set to $k_\theta = 1\,\text{N}\,\text{m}\,\text{rad}^{-1}$. Numerical accuracy was assessed by monitoring conservation of the total mechanical energy, $E_{\text{tot}}(t) = \sum_{i=1}^{N} \left[ U_i(t) + T_i^{\text{trans}}(t) + T_i^{\text{rot}}(t) \right]$, where the three terms denote the potential, translational kinetic, and rotational kinetic energies of the $i$th cell, respectively.

For the frequency analysis in Figures. 2b and 4c, a two-dimensional FFT (fast Fourier transform) was applied to $\widetilde{q}_i(t)$ from 200-cell simulations, with durations of 3 s for configurations i–iv and 5 s for configuration v. The nonuniform `ode45` responses were linearly interpolated onto 16,000 uniform time samples, and the FFT magnitude was normalized by the temporal and spatial sample counts and plotted against frequency and $\kappa$ using a common color-axis upper limit of $10^{-4}$.

To form the numerical dispersion curve in Figure 3a from the Fourier spectrum in Figure 2b, the nonnegative wavenumber range from 0 to $\pi$ was divided into 1,000 bins. Within each bin, local spectral maxima were identified, and the continuous ridge associated with the bulk dispersion was extracted. The edge-mode frequency was identified separately from the nearly horizontal spectral feature. The extracted ridge points were smoothed using moving-mean filters and interpolated onto 1,000 wavenumber points using shape-preserving cubic interpolation.

## 4.2 Fabrication and experimental setup

The unit cells were fabricated from PLA filament using a Bambu Lab X1 Carbon 3D printer. The experimental setup was assembled using commercially available bearings, linear bushings, shafts, and torsional springs. The detailed specifications, product names, and other relevant information are provided in Supplementary Note 8. Each sliding base contained a single linear bushing guided along one shaft to minimize friction.

The unit cells used for configurations i and iii had an equilibrium marker separation of approximately 7.4 cm and a mass of approximately 13.6 g. Those used for configurations ii and iv had a marker separation of approximately 3.7 cm and a mass of approximately 14.5 g, whereas configuration v had corresponding values of approximately 1.9 cm and 14.4 g.

Small variations in the total mass and mass distribution were unavoidable because the unit-cell designs differed among the configurations. The experiments were therefore intended to assess whether the numerically predicted configuration-dependent propagation and localization trends could be reproduced in a physical system, rather than to provide a parameter-fitted quantitative comparison.

## 4.3 Dynamic experiments and data processing

Dynamic experiments were conducted for configurations i–v, each with five repeated trials. The first unit cell was held at an initial generalized-coordinate displacement of 30° using a loop-shaped metal wire, which was subsequently cut to initiate the excitation. The motion was recorded using an iPhone 14 Pro positioned approximately 0.8 m above the midpoint of the shaft array, with a frame rate of 240 frames $\text{s}^{-1}$. Marker positions were determined from the centroids of the marker regions detected by HSV-based color tracking.

The representative result shown in Figure 5b was selected as the individual trial with the smallest mean-squared difference from the ensemble-averaged generalized-coordinate field over all cells and synchronized time points, and thus most closely represents the average behavior across the repeated experiments. The tracked marker trajectories were smoothed using a third-order Savitzky–Golay filter with an 11-frame window before calculating the generalized coordinates and velocities. Nearby odd window lengths produced the same qualitative IPR and normalized peak energy trends.

The generalized coordinate of each cell was reconstructed from the projected separation between markers on adjacent shafts after applying a geometric correction for the camera perspective. The characteristic length $\ell_i$ of each cell was first determined from its equilibrium marker separation $d_i^0$ using the prescribed length ratio $\Lambda$ and rest angle $\theta_0$:

$$\ell_i = \frac{d_i^0}{\sqrt{\Lambda^2 + 1 - 2\Lambda\cos\theta_0}}. \tag{17}$$

Defining the normalized instantaneous marker separation as $D_i(t) = d_i(t)/\ell_i$, the generalized coordinate was calculated using the law of cosines as

$$q_i(t) = \sin^{-1}\left[\frac{D_i^2(t) + 1 - \Lambda^2}{2D_i(t)}\right], \qquad -\frac{\pi}{2} \leq q_i(t) \leq \frac{\pi}{2}. \tag{18}$$

The velocities required for the energy calculation were subsequently obtained by numerically differentiating the smoothed trajectories. The processed displacement and velocity data were used to calculate the cell-wise kinetic and elastic potential energies, from which the inverse participation ratio and normalized peak energy were evaluated.

# Author Contributions

M.J. conceived the geometric design, performed the numerical simulations, developed the singular-edge-mode analysis and related analyses, designed and conducted the experiments, and prepared the manuscript and figures. D.K. derived the equations of motion, performed the modal analysis, and contributed to the experiments and manuscript. R.C., P.G.K., and J.Y. provided guidance on the analyses, reviewed and edited the manuscript, and supervised the research. All authors discussed the results and approved the final manuscript.

# Acknowledgments

We thank Dr. Georgios Theocharis (LAUM, France) for his valuable input. This work was supported by SNU-IAMD, SNU-IOER, the National Research Foundation of Korea grants funded by the Korean government (2023R1A2C2003705 and 2022H1D3A2A03096579), and the Korea Basic Science Institute National Research Facilities and Equipment Center grant funded by the Ministry of Science and ICT (RS-2025-00564733).

# Conflicts of Interest

The authors declare no conflicts of interest.

# Data Availability Statement

The data and numerical and analytical codes that support the findings of this study are openly available in Zenodo at https://doi.org/10.5281/zenodo.21822024.

## Supporting Information

Additional supporting information can be found online in the Supporting Information section.

## Supporting Information

# Extreme Localization via a Singular Edge Mode in an Asymmetric Linkage System

Myeonggyun Joo, Dongwook Kwak, Rajesh Chaunsali,
Panayotis G. Kevrekidis, and Jinkyu Yang*

*Corresponding author: jkyang11@snu.ac.kr

## Supplementary Text

Supplementary Note 1: Static properties

Studies of wave propagation in mechanical metamaterials often begin with the characterization and tuning of their static response, particularly the force–displacement relation. To generally parameterize the unit cell of the present system, we assign subscript 1 to the quantities associated with the shorter bar, including the generalized coordinate, and subscript 2 to those associated with the opposite bar. The static force–displacement relation is obtained by using $\theta_1$ as a continuation parameter and determining $\theta_2$ from the geometric constraint of the linkage. For the configuration branch considered here,

$$l_1 \cos\theta_1 = l_2 \cos\theta_2, \tag{S1}$$

such that

$$\theta_2(\theta_1) = \cos^{-1}\left(\frac{l_1}{l_2}\cos\theta_1\right), \quad 0 \le \theta_2 \le \frac{\pi}{2} \tag{S2}$$

where the branch of the inverse cosine is selected consistently with the prescribed geometric configuration. Depending on the geometric type of the unit cell, $l_1$ and $l_2$ correspond to $L$ and $R$, respectively, or vice versa.

The angular deformation of the torsional spring is

$$\Delta\theta = \theta_1 + \theta_2 - \bar{\theta}, \qquad \bar{\theta} = \bar{\theta_1} + \bar{\theta_2}, \tag{S3}$$

where $\bar{\theta_1}$ and $\bar{\theta_2}$ denote the equilibrium angles of each bar. The corresponding restoring moment is

$$M_s = k_\theta \Delta\theta. \tag{S4}$$

From the static force and moment balances of the two rigid bars, the horizontal force applied to the linkage is obtained as

$$F_{\text{static}} = \frac{-\dfrac{k_\theta \Delta\theta \tan\theta_1}{2l_1 \sin\theta_1} - \dfrac{k_\theta \Delta\theta \tan\theta_1}{2l_2 \sin\theta_2}}{\dfrac{\tan\theta_1}{\tan\theta_2} + 1}. \tag{S5}$$

Equivalently, Equation (S5) can be expressed in the compact form

$$F_{\text{static}} = -\frac{k_\theta \Delta\theta \,(l_1 \sin\theta_1 + l_2 \sin\theta_2)}{2 l_1 l_2 \sin(\theta_1 + \theta_2)}. \tag{S6}$$

The horizontal displacement relative to the equilibrium configuration is defined as

$$u = \left(2l_1 \sin\bar{\theta_1} + 2l_2 \sin\bar{\theta_2}\right) - \left(2l_1 \sin\theta_1 + 2l_2 \sin\theta_2\right). \tag{S7}$$

Under this convention, positive and negative values of $u$ correspond to compression and tension, respectively.

The force–displacement curves calculated from Equations. (S6) and (S7) are shown in Figure S1. For this comparison, the total bar length was scaled up by a factor of 20 relative to that used in the numerical simulations to simplify the numerical values, while the torsional stiffness was fixed at the same representative value:

$$l_1 + l_2 = 3\,\mathrm{m}, \qquad k_\theta = 1\,\mathrm{N\,m\,rad^{-1}}. \tag{S8}$$

For the symmetric systems with $\Lambda = 1$, shown in Figure S1a–c, the force–displacement curves exhibit hardening under tension and softening under compression. As the rest angle increases, the displacement range over which softening occurs becomes broader, while the overall qualitative trend remains unchanged.

In contrast, the asymmetric linkage systems shown in Figure S1d–i exhibit hardening behavior under both tension and compression. Moreover, at the same rest angle, the $L$-type system with $\Lambda = 0.5$ and the $R$-type system with $\Lambda = 2$ produce identical force–displacement curves. In particular, the mirror-related unit-cell pairs, namely configurations i and iii and configurations ii and iv, possess identical static properties despite exhibiting markedly different dynamic responses, as demonstrated in the main text. This result shows that static characterization alone cannot distinguish the configurations responsible for the observed wave propagation and localization behavior. Therefore, the dynamics of the system must be considered explicitly to capture the phenomena investigated in this study.

Supplementary Note 2: Equation of motion

***Derivation of Equation of Motion*** — The geometric constraint requires that the vertical positions of the upper joints of the two linkages within each unit cell be identical, ensuring connection of the upper joint. Accordingly, the constraint is written as

$$L_i \cos\theta_i^{\mathrm{L}} = R_i \cos\theta_i^{\mathrm{R}}. \tag{S9}$$

Taking the time derivative of the above constraint yields

$$-L_i \sin\theta_i^{\mathrm{L}}\,\dot{\theta}_i^{\mathrm{L}} = -R_i \sin\theta_i^{\mathrm{R}}\,\dot{\theta}_i^{\mathrm{R}}. \tag{S10}$$

The superscripts L and R refer to the *left* and *right bars* within each unit cell, respectively. In the present system, the generalized coordinate is defined as either $\theta^{\mathrm{R}}$ or $\theta^{\mathrm{L}}$, depending on the geometric type (*L*- or *R*-type). Therefore, using Equations. (S9) and (S10), the system can be expressed solely in terms of $\vec{q}$.

The principle of virtual power provides an effective framework for describing the dynamics of systems subject to geometric constraints. To derive the equation of motion through the principle of virtual power, we introduce the redundant coordinate vector of unit $i$, which contains all translational and rotational degrees of freedom associated with the rigid-body motions of the two bars:

$$\vec{r}_i = \begin{bmatrix} x_i^{\mathrm{L}} & y_i^{\mathrm{L}} & \theta_i^{\mathrm{L}} & x_i^{\mathrm{R}} & y_i^{\mathrm{R}} & \theta_i^{\mathrm{R}} \end{bmatrix}^T \in \mathbb{R}^6. \tag{S11}$$

The positions within each unit cell are first described relative to the lower joint of its right bar. The corresponding local coordinates of the bar centers are

$$\begin{aligned} x_{i,\mathrm{local}}^{\mathrm{L}} &= -L_i \sin\theta_i^{\mathrm{L}} - 2R_i \sin\theta_i^{\mathrm{R}}, & y_{i,\mathrm{local}}^{\mathrm{L}} &= L_i \cos\theta_i^{\mathrm{L}}, \\ x_{i,\mathrm{local}}^{\mathrm{R}} &= -R_i \sin\theta_i^{\mathrm{R}}, & y_{i,\mathrm{local}}^{\mathrm{R}} &= R_i \cos\theta_i^{\mathrm{R}}. \end{aligned} \tag{S12}$$

However, the lower joint of the right bar of unit $i$ is not fixed, except for the last unit. Its horizontal position is determined by the accumulated deformation of all unit cells located between unit $i$ and the fixed right boundary. The horizontal span of unit $j$ is

$$d_j = 2L_j \sin\theta_j^{\mathrm{L}} + 2R_j \sin\theta_j^{\mathrm{R}}, \tag{S13}$$

and the global horizontal position of the lower-right joint of unit $i$ is therefore

$$X_i = -\sum_{j=i+1}^{N} d_j, \qquad X_N = 0. \tag{S14}$$

The negative sign arises because the coordinates are accumulated from the fixed right boundary toward the left, along the negative $x$ direction. Accordingly, the global coordinates entering the equation of motion are

$$\begin{aligned} x_i^{\mathrm{L}} &= X_i - L_i \sin\theta_i^{\mathrm{L}} - 2R_i \sin\theta_i^{\mathrm{R}}, & y_i^{\mathrm{L}} &= y_{i,\mathrm{local}}^{\mathrm{L}} = L_i \cos\theta_i^{\mathrm{L}}, \\ x_i^{\mathrm{R}} &= X_i - R_i \sin\theta_i^{\mathrm{R}}, & y_i^{\mathrm{R}} &= y_{i,\mathrm{local}}^{\mathrm{R}} = R_i \cos\theta_i^{\mathrm{R}}. \end{aligned} \tag{S15}$$

The redundant coordinate vector of the complete system is then

$$\vec{r} = \begin{bmatrix} \vec{r}_1^{\,T} & \cdots & \vec{r}_N^{\,T} \end{bmatrix}^T \in \mathbb{R}^{6N}. \tag{S16}$$

Differentiating Equation (S14) gives the accumulated horizontal velocity of the lower-right joint of unit $i$:

$$\dot{X}_i = -\sum_{j=i+1}^{N} \left( 2L_j \cos\theta_j^{\mathrm{L}}\dot{\theta}_j^{\mathrm{L}} + 2R_j \cos\theta_j^{\mathrm{R}}\dot{\theta}_j^{\mathrm{R}} \right). \tag{S17}$$

The velocity components of unit $i$ are therefore

$$\vec{v}_i = \dot{\vec{r}}_i = \begin{bmatrix} \dot{x}_i^{\mathrm{L}} & \dot{y}_i^{\mathrm{L}} & \dot{\theta}_i^{\mathrm{L}} & \dot{x}_i^{\mathrm{R}} & \dot{y}_i^{\mathrm{R}} & \dot{\theta}_i^{\mathrm{R}} \end{bmatrix}^T, \tag{S18}$$

where

$$
\begin{aligned}
\dot{x}_i^{\mathrm{L}} &= \dot{X}_i - L_i \cos\theta_i^{\mathrm{L}} \dot{\theta}_i^{\mathrm{L}} - 2R_i \cos\theta_i^{\mathrm{R}} \dot{\theta}_i^{\mathrm{R}}, \\
\dot{y}_i^{\mathrm{L}} &= -L_i \sin\theta_i^{\mathrm{L}} \dot{\theta}_i^{\mathrm{L}}, \\
\dot{x}_i^{\mathrm{R}} &= \dot{X}_i - R_i \cos\theta_i^{\mathrm{R}} \dot{\theta}_i^{\mathrm{R}}, \\
\dot{y}_i^{\mathrm{R}} &= -R_i \sin\theta_i^{\mathrm{R}} \dot{\theta}_i^{\mathrm{R}}.
\end{aligned}
\tag{S19}
$$

The complete velocity vector is

$$
\vec{v} = \begin{bmatrix} \vec{v}_1^{\,T} & \cdots & \vec{v}_N^{\,T} \end{bmatrix}^T \in \mathbb{R}^{6N}. \tag{S20}
$$

Equation (S17) shows that the velocity of unit $i$ depends not only on its own generalized velocity but also on the generalized velocities of all units located between unit $i$ and the fixed right boundary. After applying the geometric constraint in Equation (S10), the two angular velocities of each unit can be expressed in terms of the single generalized velocity $\dot{q}_i$. Consequently, the redundant velocity of unit $i$ can be written as

$$
\vec{v}_i = \vec{G}_i(\vec{q})\dot{q}_i + \sum_{j=i+1}^{N} \vec{g}_j(\vec{q})\dot{q}_j, \tag{S21}
$$

where $\vec{G}_i \in \mathbb{R}^{6\times 1}$ represents the local kinematic contribution of unit $i$, whereas $\vec{g}_j \in \mathbb{R}^{6\times 1}$ represents the cumulative horizontal translation transmitted from unit $j$ to the units located to its left.

Since all velocity components are linear in the generalized velocities $\dot{\vec{q}}$, the global velocity vector can be expressed using the Jacobian matrix $\mathbf{G}(\vec{q}) \in \mathbb{R}^{6N\times N}$ as

$$
\vec{v} = \mathbf{G}(\vec{q})\dot{\vec{q}}. \tag{S22}
$$

Differentiating this relation with respect to time gives

$$
\vec{a} = \dot{\vec{v}} = \dot{\mathbf{G}}(\vec{q}, \dot{\vec{q}})\dot{\vec{q}} + \mathbf{G}(\vec{q})\ddot{\vec{q}}. \tag{S23}
$$

The Jacobian matrix $\mathbf{G}(\vec{q}) \in \mathbb{R}^{6N\times N}$ takes the following block-triangular form:

$$
\mathbf{G} = \begin{bmatrix}
\vec{G}_1 & \vec{g}_2 & \vec{g}_3 & \cdots & \vec{g}_i & \cdots & \vec{g}_N \\
\vec{0}_{6\times1} & \vec{G}_2 & \vec{g}_3 & \cdots & \vec{g}_i & \cdots & \vec{g}_N \\
\vec{0}_{6\times1} & \vec{0}_{6\times1} & \vec{G}_3 & \cdots & \vec{g}_i & \cdots & \vec{g}_N \\
\vdots & \vdots & \vdots & \ddots & \ddots & \ddots & \vdots \\
\vec{0}_{6\times1} & \vec{0}_{6\times1} & \vec{0}_{6\times1} & \cdots & \vec{G}_i & \cdots & \vec{g}_N \\
\vdots & \vdots & \vdots & \ddots & \ddots & \ddots & \vdots \\
\vec{0}_{6\times1} & \vec{0}_{6\times1} & \vec{0}_{6\times1} & \cdots & \vec{0}_{6\times1} & \cdots & \vec{G}_N
\end{bmatrix}. \tag{S24}
$$

where each diagonal block $\vec{G}_i \in \mathbb{R}^{6\times 1}$ represents the local kinematic contribution of the generalized velocity $\dot{q}_i$ to the redundant velocity of unit $i$, whereas each off-diagonal block $\vec{g}_j \in \mathbb{R}^{6\times 1}$ represents the cumulative horizontal translation induced by the generalized velocity $\dot{q}_j$. The upper block-triangular structure follows directly from the cumulative coordinate definition. In particular, the $i$-th block row of $\mathbf{G}$ shows that the redundant velocity of unit $i$ depends on its own generalized velocity $\dot{q}_i$ through $\vec{G}_i$ and on the generalized velocities $\dot{q}_j$, with $j = i+1, \ldots, N$, through $\vec{g}_j$. These latter terms arise because the motion of the units located between unit $i$ and the fixed right boundary produces a cumulative horizontal translation of unit $i$. Equivalently, when the Jacobian is interpreted columnwise, the generalized velocity $\dot{q}_i$ affects the redundant velocity of unit $i$ through $\vec{G}_i$ and those of units $1, \ldots, i-1$ through $\vec{g}_i$, but it does not affect units $i+1, \ldots, N$. Thus, the deformation of a given unit is transmitted toward the free left boundary, while the units located closer to the fixed right boundary remain unaffected. The explicit expressions for $\vec{G}_i$ and $\vec{g}_i$ depend on the geometric type of the unit cell (*L*- or *R*-type). They are obtained by substituting the velocity constraint in Equation (S10) into the velocity expressions and rewriting all angular velocities in terms of the generalized velocity $\dot{q}_i$. Their detailed forms are provided in the below section.

Each unit cell possesses two rigid linkages with distinct inertial properties. The inertia of the $i$-th cell, expressed in the redundant coordinate space, is defined as

$$\hat{\mathbf{M}}_i = \text{diag}\left[m_i^{\text{L}} \quad m_i^{\text{L}} \quad J_i^{\text{L}} \quad m_i^{\text{R}} \quad m_i^{\text{R}} \quad J_i^{\text{R}}\right], \tag{S25}$$

$$\hat{\mathbf{M}} = \text{blkdiag}(\hat{\mathbf{M}}_1, \ldots, \hat{\mathbf{M}}_N). \tag{S26}$$

Here, diag$(\cdot)$ denotes a diagonal matrix, and blkdiag$(\cdot)$ denotes a block-diagonal matrix constructed from the listed matrix blocks. The quantities $m_i^{\text{L}}$, $m_i^{\text{R}}$ and $J_i^{\text{L}}$, $J_i^{\text{R}}$ denote the mass and rotational inertia of the left and right linkages, respectively.

Meanwhile, let $\bar{\theta}_i^{\text{L}}$ and $\bar{\theta}_i^{\text{R}}$ be the equilibrium (rest) angles of the left and right linkages measured from the vertical direction. The equilibrium configuration of the torsional spring is defined by their sum

$$\bar{\theta}_i = \bar{\theta}_i^{\text{L}} + \bar{\theta}_i^{\text{R}}, \tag{S27}$$

so that the spring produces no moment when $\theta_i^{\text{L}} + \theta_i^{\text{R}} = \bar{\theta}_i$. This quantity represents the intrinsic geometric angle formed between the left and right bars of the unit cell. Deviations from this equilibrium generate equal and opposite torques on the two linkages proportional to $(\theta_i^{\text{L}} + \theta_i^{\text{R}} - \bar{\theta}_i)$.

We now consider the entire system. At the left free end ($i = 1$), an external horizontal force $F_{ex}$ can be applied to the lower slider, while experiencing the torsional spring moment. The remaining units ($i = 2, \ldots, N$) experience only the torsional spring moments. In the ideal model, gravitational and damping effects are neglected. Under these assumptions, the virtual power balance of the system considering virtual velocity perturbation can be written as

$$\sum_{i=1}^{N} \left(\hat{\mathbf{M}}_i \, \dot{\vec{v}}_i - \vec{f}_i\right) \cdot \delta\vec{v}_i = 0, \tag{S28}$$

where $\vec{f}_i$ represents the force vector of the $i$-th unit in the redundant coordinate space:

$$\vec{f}_i = \begin{cases} \left[F_{ex} \quad 0 \quad -F_{ex}L_1\cos(\theta_1^{\text{L}}) - k_\theta(\theta_1^{\text{L}} + \theta_1^{\text{R}} - \bar{\theta}_1) \quad 0 \quad 0 \quad -k_\theta(\theta_1^{\text{L}} + \theta_1^{\text{R}} - \bar{\theta}_1)\right]^T, & (i = 1), \\ \left[0 \quad 0 \quad -k_\theta(\theta_i^{\text{L}} + \theta_i^{\text{R}} - \bar{\theta}_i) \quad 0 \quad 0 \quad -k_\theta(\theta_i^{\text{L}} + \theta_i^{\text{R}} - \bar{\theta}_i)\right]^T, & (i = 2, \ldots, N). \end{cases} \tag{S29}$$

Here, the physical meaning of each component in $\vec{f}_i$ depends on the unit index. For the free-end unit ($i = 1$), the first component represents the horizontal force acting on the left linkage due to the external excitation $F_{ex}$, which is present only under the external-force condition described in Supplementary Note 3. When $F_{ex}$ is applied, the third component combines two contributions: the external moment generated by $F_{ex}$ through the lever arm $L_1 \cos\theta_1^{\text{L}}$ and the torsional restoring moment of the upper spring. The sixth component corresponds solely to the torsional spring moment acting on the right linkage. The remaining components of $\vec{f}_1$ are zero because no vertical and translational forces acting on the right linkage are directly applied to the system. For the interior and fixed-end units ($i = 2, \ldots, N$), only the torsional springs contribute to the generalized forces. In this case, the third and sixth components represent the spring moments acting on the left and right linkages, respectively, while all other components vanish. This formulation enforces that, at every instant, the total virtual power of inertia and external forces is balanced, forming the basis for the global equation of motion.

Using $\delta\vec{v} = \mathbf{G}\,\delta\dot{\vec{q}}$ and the total stacked forms $\hat{\mathbf{M}} = \text{blkdiag}(\hat{\mathbf{M}}_1, \ldots, \hat{\mathbf{M}}_N)$, $\vec{f} = \left[\vec{f}_1^T \quad \cdots \quad \vec{f}_N^T\right]^T$, we obtain

$$\mathbf{G}^T\hat{\mathbf{M}}\,\dot{\vec{v}} = \mathbf{G}^T\vec{f}, \qquad \dot{\vec{v}} = \dot{\mathbf{G}}\,\dot{\vec{q}} + \mathbf{G}\,\ddot{\vec{q}}. \tag{S30}$$

At this stage, the equation is expressed in a form premultiplied by $\mathbf{G}^T$, which serves as a preliminary step toward ensuring the symmetry of the system matrix derived subsequently. This formulation improves the efficiency of the numerical simulations and provides the basis for the subsequent modal analysis. The final equation of motion expressed in terms of the generalized coordinates is

$$\mathbf{G}^T\hat{\mathbf{M}}\mathbf{G}\,\ddot{\vec{q}} + \mathbf{G}^T\hat{\mathbf{M}}\dot{\mathbf{G}}\,\dot{\vec{q}} = \mathbf{G}^T\vec{f}. \tag{S31}$$

***Jacobian*** — The explicit forms of the Jacobian sub-vectors $\vec{G}_i$ and $\vec{g}_i$ depend on the geometric configuration of each unit cell. They are obtained by separating the contributions associated with each generalized coordinate. For compactness, let $s^{\mathrm{L}} = \sin\theta_i^{\mathrm{L}}$, $s^{\mathrm{R}} = \sin\theta_i^{\mathrm{R}}$, $c^{\mathrm{L}} = \cos\theta_i^{\mathrm{L}}$, and $c^{\mathrm{R}} = \cos\theta_i^{\mathrm{R}}$.

For the L-type configuration ($q_i = \theta_i^{\mathrm{R}}$), the local Jacobian $\vec{G}_i$ and the propagation term $\vec{g}_i$ are given by

$$\vec{G}_i = \begin{bmatrix} G_{i1} & G_{i2} & G_{i3} & G_{i4} & G_{i5} & G_{i6} \end{bmatrix}^T, \tag{S32}$$

$$\begin{aligned} G_{i1} &= -R\frac{s^{\mathrm{R}}}{s^{\mathrm{L}}}c^{\mathrm{L}} - 2Rc^{\mathrm{R}}, \qquad G_{i2} = -Rs^{\mathrm{R}}, \qquad G_{i3} = \frac{R}{L}\frac{s^{\mathrm{R}}}{s^{\mathrm{L}}}, \\ G_{i4} &= -Rc^{\mathrm{R}}, \qquad G_{i5} = -Rs^{\mathrm{R}}, \qquad G_{i6} = 1, \end{aligned} \tag{S33}$$

$$\vec{g}_i = \begin{bmatrix} g_{i1} & 0 & 0 & g_{i4} & 0 & 0 \end{bmatrix}^T, \qquad g_{i1} = g_{i4} = -2R\frac{s^{\mathrm{R}}}{s^{\mathrm{L}}}c^{\mathrm{L}} - 2Rc^{\mathrm{R}}. \tag{S34}$$

For the R-type configuration ($q_i = \theta_i^{\mathrm{L}}$), the corresponding terms are expressed as

$$\vec{G}_i = \begin{bmatrix} G_{i1} & G_{i2} & G_{i3} & G_{i4} & G_{i5} & G_{i6} \end{bmatrix}^T, \tag{S35}$$

$$\begin{aligned} G_{i1} &= -2L\frac{s^{\mathrm{L}}}{s^{\mathrm{R}}}c^{\mathrm{R}} - Lc^{\mathrm{L}}, \qquad G_{i2} = -Ls^{\mathrm{L}}, \qquad G_{i3} = 1, \\ G_{i4} &= -L\frac{s^{\mathrm{L}}}{s^{\mathrm{R}}}c^{\mathrm{R}}, \qquad G_{i5} = -Ls^{\mathrm{L}}, \qquad G_{i6} = \frac{L}{R}\frac{s^{\mathrm{L}}}{s^{\mathrm{R}}}, \end{aligned} \tag{S36}$$

$$\vec{g}_i = \begin{bmatrix} g_{i1} & 0 & 0 & g_{i4} & 0 & 0 \end{bmatrix}^T, \qquad g_{i1} = g_{i4} = -2L\frac{s^{\mathrm{L}}}{s^{\mathrm{R}}}c^{\mathrm{R}} - 2Lc^{\mathrm{L}}. \tag{S37}$$

Using these expressions, the Jacobian terms can be substituted into the following relations to evaluate the $i$-th diagonal component $\vec{G}_i^T\hat{\mathbf{M}}_i\vec{G}_i$, the subsequent diagonal component $\vec{g}_i^T\hat{\mathbf{M}}_i\vec{G}_j$, and the off-diagonal component $\vec{g}_i^T\hat{\mathbf{M}}_i\vec{g}_i$ as

$$\vec{G}_i^T\hat{\mathbf{M}}_i\vec{G}_i = \left(G_{i1}^2 + G_{i2}^2\right)m_i^{\mathrm{L}} + \left(G_{i4}^2 + G_{i5}^2\right)m_i^{\mathrm{R}} + G_{i3}^2 J_i^{\mathrm{L}} + G_{i6}^2 J_i^{\mathrm{R}}, \tag{S38}$$

$$\vec{G}_i^T\hat{\mathbf{M}}_i\vec{g}_j = \vec{g}_j^T\hat{\mathbf{M}}_i\vec{G}_i = g_{j1}G_{i1}m_i^{\mathrm{L}} + g_{j4}G_{i4}m_i^{\mathrm{R}}, \qquad j > i, \tag{S39}$$

$$\vec{g}_j^T\hat{\mathbf{M}}_i\vec{g}_k = g_{j1}g_{k1}m_i^{\mathrm{L}} + g_{j4}g_{k4}m_i^{\mathrm{R}}, \qquad j, k > i. \tag{S40}$$

***System Matrix*** — As mentioned in the main text, the complete governing equation of motion and linearized equation of motion can be written as follows,

$$\begin{aligned} &\mathcal{M}(\vec{q})\ddot{\vec{q}} + \mathbf{C}(\vec{q},\dot{\vec{q}})\dot{\vec{q}} - \mathbf{G}^T\vec{f} = \vec{0}, \\ &\rightarrow \bar{\mathcal{M}}\ddot{\vec{q}} + \bar{\mathbf{K}}\vec{q} = \vec{0}. \end{aligned} \tag{S41}$$

*Mass Matrix* — We now provide a more detailed examination of the system matrices that emerge during the derivation process. The mass matrix can be derived both before and after linearization, with the linearized mass matrix $\bar{\mathcal{M}}$ obtained by evaluating the configuration-dependent matrix $\mathcal{M}$ introduced below at the equilibrium configuration $\vec{q}_0$. The global mass matrix $\mathcal{M}$ is constructed by assembling the per-unit inertia contributions through the kinematic mapping $\mathbf{G}$:

$$\begin{aligned} \mathcal{M} &= \mathbf{G}^T\hat{\mathbf{M}}\mathbf{G} \\ &= \left[\begin{array}{c:ccc} \vec{G}_1^T\hat{\mathbf{M}}_1\vec{G}_1 & \vec{G}_1^T\hat{\mathbf{M}}_1\vec{g}_2 & \cdots\cdots & \vec{G}_1^T\hat{\mathbf{M}}_1\vec{g}_N \\ \hdashline \vec{g}_2^T\hat{\mathbf{M}}_1\vec{G}_1 & \vec{g}_2^T\hat{\mathbf{M}}_1\vec{g}_2 & \cdots\cdots & \vec{g}_2^T\hat{\mathbf{M}}_1\vec{g}_N \\ \vdots & \vdots & \ddots & \vdots \\ \vec{g}_N^T\hat{\mathbf{M}}_1\vec{G}_1 & \vec{g}_N^T\hat{\mathbf{M}}_1\vec{g}_2 & \cdots\cdots & \vec{g}_N^T\hat{\mathbf{M}}_1\vec{g}_N \end{array}\right] \end{aligned}$$

$$+\cdots+\begin{bmatrix}0 & \cdots & 0\\ \vdots & \ddots & \vdots\\ 0 & \cdots & \vec{G}_N^T\hat{\mathbf{M}}_N\vec{G}_N\end{bmatrix} \tag{S42}$$

$$=\sum_{i=1}^{N}\mathcal{M}_i. \tag{S43}$$

$$\mathcal{M}_i=\begin{bmatrix}0 & \cdots & \cdots & \cdots & 0\\ \vdots & \vec{G}_i^T\hat{\mathbf{M}}_i\vec{G}_i & \vec{G}_i^T\hat{\mathbf{M}}_i\vec{g}_{i+1} & \cdots & \vec{G}_i^T\hat{\mathbf{M}}_i\vec{g}_N\\ \vdots & \vec{g}_{i+1}^T\hat{\mathbf{M}}_i\vec{G}_i & \vec{g}_{i+1}^T\hat{\mathbf{M}}_i\vec{g}_{i+1} & \cdots & \vec{g}_{i+1}^T\hat{\mathbf{M}}_i\vec{g}_N\\ \vdots & \vdots & \vdots & \ddots & \vdots\\ 0 & \vec{g}_N^T\hat{\mathbf{M}}_i\vec{G}_i & \vec{g}_N^T\hat{\mathbf{M}}_i\vec{g}_{i+1} & \cdots & \vec{g}_N^T\hat{\mathbf{M}}_i\vec{g}_N\end{bmatrix}. \tag{S44}$$

Here, each submatrix $\mathcal{M}_i$ corresponds to the contribution of the $i$-th unit cell. Because of the cumulative kinematic constraint imposed by the fixed–free boundary condition, the motion of unit $i$ contributes not only to its own matrix entry but also to the entries associated with units $1,\ldots,i-1$, which are located between unit $i$ and the free left boundary.

For a uniform (homogeneous) system composed of identical unit cells, the unit index $i$ may be omitted, and the corresponding quantities can be written as $\vec{G}_i=\vec{G}$, $\vec{g}_i=\vec{g}$, and $\hat{\mathbf{M}}_i=\hat{\mathbf{M}}$. The three scalar inertial coefficients are then defined as

$$\mathcal{G}=\vec{G}^T\hat{\mathbf{M}}\vec{G}, \tag{S45}$$

$$\gamma=\vec{G}^T\hat{\mathbf{M}}\vec{g}, \tag{S46}$$

$$\mathfrak{g}=\vec{g}^T\hat{\mathbf{M}}\vec{g}. \tag{S47}$$

Here, $\mathcal{G}$, $\gamma$, and $\mathfrak{g}$ are scalar inertial coefficients determined by the unit-cell geometry and inertial properties. These quantities can be calculated from Equations. (S38)–(S40), and their explicit forms are provided in Table 1. Under this uniform assumption, the assembled mass matrix becomes

$$\mathcal{M}_U=\begin{bmatrix}\mathcal{G} & \gamma & \cdots & \gamma\\ \gamma & \mathfrak{g} & \cdots & \mathfrak{g}\\ \vdots & \vdots & \ddots & \vdots\\ \gamma & \mathfrak{g} & \cdots & \mathfrak{g}\end{bmatrix}+\begin{bmatrix}0 & 0 & \cdots & 0\\ 0 & \mathcal{G} & \cdots & \gamma\\ \vdots & \vdots & \ddots & \vdots\\ 0 & \gamma & \cdots & \mathfrak{g}\end{bmatrix}+\cdots+\begin{bmatrix}0 & \cdots & 0\\ \vdots & \ddots & \vdots\\ 0 & \cdots & \mathcal{G}\end{bmatrix}. \tag{S48}$$

In Equations. (S42), (S44), and (S48), the entries indicated beyond the dotted guides represent the terms that capture the cumulative inertial influence transmitted toward the free left boundary. Therefore, unlike a conventional mass–stiffness system whose mass matrix typically takes a diagonal or narrowly banded form due to local interactions, the present system exhibits a **fully populated matrix** arising from cumulative inertial coupling. Hence, a coordinate transformation is required to facilitate further modal analysis.

*Stiffness Matrix*— The linearized stiffness matrix $\bar{\mathbf{K}}$ is obtained by linearizing the restoring-force term $-\mathbf{G}^T\vec{f}$ about the equilibrium configuration $\vec{q}_0$, with $F_{\text{ex}}=0$:

$$\bar{\mathbf{K}}=-\left.\frac{\partial\left(\mathbf{G}^T\vec{f}\right)}{\partial\vec{q}}\right|_{\vec{q}=\vec{q}_0}. \tag{S49}$$

Because the torsional-spring force vanishes at equilibrium, $\vec{f}(\vec{q}_0)=\vec{0}$, the contribution involving the derivative of $\mathbf{G}^T$ vanishes at first order. Therefore,

$$\bar{\mathbf{K}}=\mathbf{G}_0^T\mathbf{K}_c,\qquad \mathbf{G}_0=\mathbf{G}(\vec{q}_0), \tag{S50}$$

where

$$\mathbf{K}_c = \text{blkdiag}\left(\vec{K}_{c1}, \ldots, \vec{K}_{cN}\right) \in \mathbb{R}^{6N\times N} \tag{S51}$$

collects the derivatives of the per-unit torsional restoring forces with respect to the generalized coordinates. The per-unit stiffness vectors $\vec{K}_{ci} \in \mathbb{R}^6$ depend on the geometric type:

$$\vec{K}_{ci} = \begin{cases} \left[0,\ 0,\ k_{\theta,i}\left(1 + \dfrac{L}{R}\dfrac{\bar{s}^{\mathrm{L}}}{\bar{s}^{\mathrm{R}}}\right),\ 0,\ 0,\ k_{\theta,i}\left(1 + \dfrac{L}{R}\dfrac{\bar{s}^{\mathrm{L}}}{\bar{s}^{\mathrm{R}}}\right)\right]^T, & (R\text{-type}), \\ \left[0,\ 0,\ k_{\theta,i}\left(1 + \dfrac{R}{L}\dfrac{\bar{s}^{\mathrm{R}}}{\bar{s}^{\mathrm{L}}}\right),\ 0,\ 0,\ k_{\theta,i}\left(1 + \dfrac{R}{L}\dfrac{\bar{s}^{\mathrm{R}}}{\bar{s}^{\mathrm{L}}}\right)\right]^T, & (L\text{-type}), \end{cases} \tag{S52}$$

where $\bar{s}^{\mathrm{L}} = \sin\bar{\theta}^{\mathrm{L}}$ and $\bar{s}^{\mathrm{R}} = \sin\bar{\theta}^{\mathrm{R}}$.

Consequently, the assembled linearized stiffness matrix is

$$\bar{\mathbf{K}} = \text{diag}\left(\vec{G}_{0,1}^T\vec{K}_{c1}, \ldots, \vec{G}_{0,N}^T\vec{K}_{cN}\right) = \text{diag}\left(k_{c1}, \ldots, k_{cN}\right), \tag{S53}$$

where $\vec{G}_{0,i}$ denotes the local Jacobian vector evaluated at the equilibrium configuration.

For a uniform system composed of identical unit cells, $k_c = k_{c1} = \cdots = k_{cN}$, and therefore

$$\bar{\mathbf{K}} = k_c\mathbf{I}, \tag{S54}$$

where $\mathbf{I} \in \mathbb{R}^{N\times N}$ is the identity matrix.

In contrast to the fully populated mass matrix generated by cumulative inertial coupling, the linearized stiffness matrix $\bar{\mathbf{K}}$ is diagonal because each torsional spring contributes locally to its corresponding generalized coordinate. The subsequent coordinate transformation converts this diagonal matrix into the tridiagonal transformed stiffness matrix introduced in Supplementary Note 4.

*Velocity-Dependent Inertial Matrix*—

$$\mathbf{C}(\vec{q}, \dot{\vec{q}}) = \mathbf{G}^T\hat{\mathbf{M}}\dot{\mathbf{G}}, \qquad \vec{c}(\vec{q}, \dot{\vec{q}}) = \mathbf{C}(\vec{q}, \dot{\vec{q}})\dot{\vec{q}}. \tag{S55}$$

For the $R$-type ($q_i = \tilde{\theta}_i^{\mathrm{L}}$) and $L$-type ($q_i = \tilde{\theta}_i^{\mathrm{R}}$) cases, the nonlinear expressions for $\dot{\vec{G}}_i$ and $\dot{\vec{g}}_i$ are obtained by differentiating the corresponding Jacobian sub-vectors. The resulting velocity-dependent inertial terms account for the Coriolis- and centrifugal-type effects arising from the configuration-dependent kinematics of the system. The velocity-dependent inertial force $\vec{c}(\vec{q}, \dot{\vec{q}})$, introduced by collecting these terms to cast the equation of motion into standard system-matrix form, is at least quadratic in $\dot{\vec{q}}$ and is therefore neglected in the linearized modal analysis.

Supplementary Note 3: Additional information on the numerical simulations

***Simulations with Other Input Conditions*** — To initiate the wave propagation response of the system, the transient input conditions can be broadly classified into three types: (1) initial-displacement, (2) initial-velocity, and (3) external-force. Since the same system can exhibit somewhat different responses depending on how the input is applied, we performed simulations under various input conditions to examine whether the trends in energy propagation and localization are preserved. Although all input conditions exhibited the same overall qualitative trends, the main text focuses on the (2) initial-velocity condition because it most clearly reveals the differences among the configurations. Specifically, an initial rotational velocity of $-18\,\mathrm{rad\,s^{-1}}$ is imposed on the first unit of a 20-cell system. As shown in Figure 2 of the main text, the degree of energy propagation and localization differs markedly depending on the geometrical characteristics of configurations i–iv.

For the (1) initial-displacement condition shown in Figure S2, a pre-compression of $-15°$ is imposed on the first cell and then released. This pre-compression angle was selected to produce a bulk-propagation amplitude comparable to the approximately $|\widetilde{q}| \approx 6°$ response observed under the (2) initial-velocity condition. The same criterion was used to determine the excitation magnitude for the (3) external-force condition described below. The overall trend is nearly identical to that obtained under the initial-velocity input. To achieve a comparable level of energy propagation in terms of wave amplitude to the other input conditions, a large initial displacement must be imposed on the first cell, which causes the system to operate in the nonlinear regime. Nevertheless, the fact that the overall trend appearing in Figure S2 remains similar indicates that the proposed system operates relatively robustly in terms of energy transmission.

For the (3) external-force condition shown in Figure S3, a rectangular force pulse of 32 N is applied to the first cell for 1 ms. The relative trend among the configurations is still maintained; however, the overall degree of wave transmission is higher than that observed under the other input conditions. The Fourier domain results in Figure S3b show that the external-force input condition induces more pronounced bulk wave transmission than the other input conditions, particularly in the R-type systems (iii and iv). This enhanced spatial spreading of energy is consequently reflected in the reduced IPR values shown in Figure S3c.

This difference arises because an external force directly enters the governing equation in Equation (S31), and the resulting motion is newly determined through the full stiffness and mass matrices of the system. Therefore, even when the force is applied only to the first cell, the response can more readily excite multiple modes and propagate through the entire structure. In contrast, initial-velocity and initial-displacement inputs do not contain external forcing terms during the subsequent free response. In these cases, the dynamics are purely governed by the modal superposition determined by the initial condition, and the contribution of each mode depends on the projection of the imposed input onto the corresponding mode shape. Thus, when an edge mode exists and the input is applied at the edge where the mode is concentrated, the response is dominated by that edge mode, leading to strong energy localization.

***Conditions for the simulations*** — As described in the Methods, all simulations were performed in MATLAB by directly integrating the nonlinear equation of motion using the widely used adaptive Runge–Kutta solver `ode45`. The solver tolerances were selected according to the input condition. For the initial-velocity simulations presented in the main text and the initial-displacement simulations presented in Supplementary Notes 3 and 7, the tolerances were set to `RelTol` $= 10^{-10}$ and `AbsTol` $= 10^{-12}$. For the external step-force simulations, tighter tolerances of `RelTol` $= 10^{-12}$ and `AbsTol` $= 10^{-14}$ were used to accurately resolve the abrupt onset and termination of the applied force (see Supplementary Notes 3 and 7). Looser tolerances were found to produce numerical instability near the discontinuities of the step loading, particularly when the force was removed; the tighter settings were therefore adopted to minimize these numerical artifacts.

The simulation duration was set to 0.3 s, which was sufficiently long for the bulk response to reach the distal end of the chain. For configuration v, the duration was extended to 0.5 s because of its relatively slow bulk response.

Supplementary Note 4: Additional information on the modal analysis

***Components of the transformed system matrices*** — Building on the formulation presented in the main text and Supplementary Note 2, the transformed mass matrix can be written as

$$\tilde{\mathcal{M}} = \begin{bmatrix} \mathcal{G} & \vec{u}^T \\ \vec{u} & \mathbf{T}_M \end{bmatrix}, \quad \mathbf{T}_M = \mathrm{tridiag}(B, A, B), \quad \vec{u} = [B, 0, \ldots, 0]^T, \tag{S56}$$

$$A = 2\mathcal{G} - 2\gamma + \mathfrak{g}, \quad B = -\mathcal{G} + \gamma. \tag{S57}$$

The transformed stiffness matrix is

$$\tilde{\mathbf{K}} = k_c \begin{bmatrix} 1 & \vec{v}^T \\ \vec{v} & \mathbf{T}_K \end{bmatrix}, \qquad \mathbf{T}_K = \mathrm{tridiag}(-1, 2, -1), \qquad \vec{v} = [-1, 0, \ldots, 0]^T. \tag{S58}$$

Here, $\mathrm{tridiag}(X, Y, Z)$ denotes a tridiagonal matrix whose main-diagonal entries are $Y$, lower off-diagonal entries are $X$, and upper off-diagonal entries are $Z$. Both transformed matrices therefore possess the same nearest-neighbor sparsity pattern, except for their first diagonal entries. The coefficient $A$ represents the diagonal bulk inertia, whereas $B$ represents the nearest-neighbor inertial coupling. The coefficient $k_c$ is the effective stiffness incorporating the torsional stiffness $k_\theta$. The quantities $\mathcal{G}$, $\gamma$, and $\mathfrak{g}$ are geometry-dependent inertial coefficients defined in Table 1 of the main text.

The transformed equation of motion can consequently be classified into three forms according to the cell index. For the first cell,

$$k_c(\phi_1 - \phi_2) + \mathcal{G}\ddot{\phi}_1 + B\ddot{\phi}_2 = 0. \tag{S59}$$

For the interior cells,

$$k_c(-\phi_{i-1} + 2\phi_i - \phi_{i+1}) + B\ddot{\phi}_{i-1} + A\ddot{\phi}_i + B\ddot{\phi}_{i+1} = 0, \qquad i = 2, \ldots, N-1. \tag{S60}$$

For the last cell,

$$k_c(-\phi_{N-1} + 2\phi_N) + B\ddot{\phi}_{N-1} + A\ddot{\phi}_N = 0. \tag{S61}$$

***Comparison between the L-type and R-type systems*** — Because the cumulative kinematic structure is common to both cell types, the $L$-type and $R$-type configurations produce transformed matrices having the same structural form. For the $L$-type configuration, the transformed mass matrix is written as

$$\tilde{\mathcal{M}}_L = \begin{bmatrix} \mathcal{G}_L & \vec{u}_L^T \\ \vec{u}_L & \mathbf{T}_{M,L} \end{bmatrix}, \quad \mathbf{T}_{M,L} = \mathrm{tridiag}(B_L, A_L, B_L), \quad \vec{u}_L = [B_L, 0, \ldots, 0]^T,$$

$$A_L = 2\mathcal{G}_L - 2\gamma_L + \mathfrak{g}_L, \qquad B_L = -\mathcal{G}_L + \gamma_L. \tag{S62}$$

Similarly, for the $R$-type configuration,

$$\tilde{\mathcal{M}}_R = \begin{bmatrix} \mathcal{G}_R & \vec{u}_R^T \\ \vec{u}_R & \mathbf{T}_{M,R} \end{bmatrix}, \quad \mathbf{T}_{M,R} = \mathrm{tridiag}(B_R, A_R, B_R), \vec{u}_R = [B_R, 0, \ldots, 0]^T, \tag{S63}$$

$$A_R = 2\mathcal{G}_R - 2\gamma_R + \mathfrak{g}_R, \qquad B_R = -\mathcal{G}_R + \gamma_R. \tag{S64}$$

For mirror-symmetric unit-cell pairs, namely configurations i and iii and configurations ii and iv, substitution of the type-dependent coefficients in Table 1 of the main text gives

$$B_L = B_R, \qquad\qquad \mathfrak{g}_L = \mathfrak{g}_R. \tag{S65}$$

Because

$$A_L = -2B_L + \mathfrak{g}_L, \qquad\qquad A_R = -2B_R + \mathfrak{g}_R, \tag{S66}$$

these relations also imply

$$A_L = A_R. \tag{S67}$$

Consequently, the off-diagonal vectors and bulk tridiagonal blocks are identical:

$$\vec{u}_L = \vec{u}_R = \vec{u}, \qquad \mathbf{T}_{M,L} = \mathbf{T}_{M,R} = \mathbf{T}_M. \tag{S68}$$

The transformed mass matrices of each mirror-symmetric pair can therefore be expressed as

$$\tilde{\mathcal{M}}_L = \begin{bmatrix} \mathcal{G}_L & \vec{u}^T \\ \vec{u} & \mathbf{T}_M \end{bmatrix}, \qquad \tilde{\mathcal{M}}_R = \begin{bmatrix} \mathcal{G}_R & \vec{u}^T \\ \vec{u} & \mathbf{T}_M \end{bmatrix}. \tag{S69}$$

Thus, the transformed mass matrices of the $L$-type and $R$-type configurations are identical in all entries except for the first diagonal entry, which is given by $\mathcal{G}_L$ and $\mathcal{G}_R$, respectively.

For the same torsional stiffness, the mirror-symmetric pairs also have identical effective stiffness coefficients and therefore satisfy

$$\tilde{\mathbf{K}}_L = \tilde{\mathbf{K}}_R. \tag{S70}$$

The distinct boundary modal behavior of the two cell types therefore originates from a difference in only the first diagonal entry of the transformed mass matrix, while their bulk mass-matrix blocks and transformed stiffness matrices remain identical.

***Characteristic equation and spatial roots*** — A spatiotemporally separable modal solution is assumed in the form

$$\begin{aligned} \phi_n(i,t) &= a_n(i)e^{j\omega_n t}, \\ a_n(i) &= P_n r_n^{i-1} + Q_n s_n^{i-1}. \end{aligned} \tag{S71}$$

Here, $a_n(i)$ denotes the spatial amplitude of the $n$th mode at the $i$th transformed coordinate. The quantities $r_n$ and $s_n$ are the two spatial factors, whereas $P_n$ and $Q_n$ are their corresponding coefficients.

Substitution of Equation (S71) into the interior equation, Equation (S60), gives the characteristic equation

$$\left(k_c + B\omega_n^2\right) z^2 - \left(2k_c - A\omega_n^2\right) z + \left(k_c + B\omega_n^2\right) = 0, \tag{S72}$$

where $z$ denotes either $r_n$ or $s_n$. The two characteristic roots are therefore

$$r_n, s_n = \frac{2k_c - A\omega_n^2 \pm \sqrt{\left(2k_c - A\omega_n^2\right)^2 - 4\left(k_c + B\omega_n^2\right)^2}}{2\left(k_c + B\omega_n^2\right)}. \tag{S73}$$

The roots satisfy

$$r_n s_n = 1. \tag{S74}$$

When the discriminant in Equation (S73) is negative, the roots form a complex-conjugate pair on the unit circle. They can then be written as

$$r_n = e^{-j\kappa_n}, \qquad s_n = e^{j\kappa_n}, \tag{S75}$$

where $\kappa_n$ is the wavenumber of the $n$th bulk mode. The corresponding spatial profile is a sinusoidal superposition of two counter-propagating components.

When the discriminant is positive, the roots are real and reciprocal. Assuming $|r_n| < 1$ and $|s_n| > 1$, the component proportional to $r_n^{i-1}$ decays away from the left boundary, whereas the component proportional to $s_n^{i-1}$ grows with the cell index. For a semi-infinite chain, boundedness requires $Q_n = 0$, leaving a purely decaying edge-mode solution. For a finite chain, both spatial components are generally retained to satisfy the boundary condition at the opposite end. However, when the chain is sufficiently long relative to the localization length, the coefficient of the growing component becomes exponentially small. The purely decaying solution therefore becomes increasingly accurate as $|r_n|$ decreases.

***Closed-form bulk-mode solution*** — The bulk-mode solution can be written as

$$\phi_n(i,t) = \cos(\kappa_n i + \beta_n) e^{j\omega_n t}, \tag{S76}$$

where $\beta_n$ is a phase offset resulting from the generally unequal coefficients of the two counter-propagating components.

Substitution of Equation (S76) into the interior equation gives the bulk-mode frequency–wavenumber relation

$$\omega_n^2 = \frac{4k_c \sin^2(\kappa_n/2)}{A + 2B\cos\kappa_n}. \tag{S77}$$

Substitution into the first boundary equation gives

$$\cos\beta_n = \frac{k_c - \omega_n^2(\mathcal{G} - 2\gamma + \mathfrak{g})}{k_c + \omega_n^2(-\mathcal{G} + \gamma)} \cos(\kappa_n + \beta_n). \tag{S78}$$

Substitution into the last boundary equation gives

$$(N+1)\kappa_n + \beta_n = \frac{2M+1}{2}\pi, \qquad M \in \mathbb{Z}, \qquad 0 < \kappa_n < \pi. \tag{S79}$$

Each admissible value of $M$ yields one bulk-mode solution. Solving Equations. (S77)–(S79) determines $\omega_n$, $\kappa_n$, and $\beta_n$ for every bulk mode of the finite chain. The corresponding spatial mode shape then follows directly from Equation (S76).

***Closed-form edge-mode solution*** — The decaying edge-mode component is written as

$$\phi_\mathrm{e}(i,t) = r_\mathrm{e}^{i-1} e^{j\omega_\mathrm{e} t}. \tag{S80}$$

The subscript e denotes quantities associated with the edge mode.

Substitution of Equation (S80) into the first boundary equation gives the edge-mode frequency

$$\omega_\mathrm{e}^2 = \frac{k_c \mathfrak{g}}{\mathcal{G}\mathfrak{g} - \gamma^2}. \tag{S81}$$

Substitution into the interior equation gives the spatial decay ratio

$$r_\mathrm{e} = \frac{\gamma}{\gamma - \mathfrak{g}}. \tag{S82}$$

Equations (S81) and (S82) directly determine the frequency and dominant spatial profile of the edge-localized mode. A negative value of $r_\mathrm{e}$ produces staggered decay, whereas a positive value produces non-staggered decay. For a sufficiently long finite chain and a strongly localized mode, the correction associated with the reciprocal growing root is exponentially small.

***Evanescent continuation and energy localization length*** — The spatial attenuation of an edge mode can also be quantified by extending the bulk-mode dispersion relation into the evanescent regime. The complex wavenumber is decomposed as

$$\kappa = \kappa_\mathrm{real} + j\kappa_\mathrm{imag}. \tag{S83}$$

Within the first Brillouin zone, $\kappa_\mathrm{real}$ is either 0 or $\pi$, corresponding to non-staggered and staggered decay, respectively. It follows that

$$\cos\kappa = \begin{cases} \cos(j\kappa_\mathrm{imag}), & \kappa_\mathrm{real} = 0, \\ \cos(\pi + j\kappa_\mathrm{imag}), & \kappa_\mathrm{real} = \pi, \end{cases}$$

$$= \pm \cosh \kappa_{\text{imag}}. \tag{S84}$$

Rearranging the bulk dispersion relation gives

$$\cos \kappa = \frac{2k_c - A\omega^2}{2k_c + 2B\omega^2}. \tag{S85}$$

The imaginary part of the wavenumber is therefore

$$\kappa_{\text{imag}} = \operatorname{arccosh} \left| \frac{2k_c - A\omega^2}{2k_c + 2B\omega^2} \right|. \tag{S86}$$

Because the modal energy scales with the square of the displacement amplitude, the corresponding energy localization length is defined as

$$L_{\text{energy}} = \frac{1}{2\kappa_{\text{imag}}}. \tag{S87}$$

A larger value of $\kappa_{\text{imag}}$ therefore corresponds to a shorter energy localization length and stronger spatial localization.

For configurations i and ii, no eigenfrequency intersects the evanescent branch, and therefore no physical spatially decaying mode occurs. For configurations iii and iv, the edge-mode frequencies intersect the corresponding evanescent branches at finite values of $\kappa_{\text{imag}}$. The intersection for configuration iv occurs at a substantially larger value of $\kappa_{\text{imag}}$ than that for configuration iii. This result is consistent with the shorter localization length and more strongly confined mode shape of configuration iv.

***Simulations with sinusoidal input*** — A chain containing $N$ unit cells possesses $N$ modal frequencies and corresponding mode shapes. The analytical framework described above provides closed-form relations for these modal characteristics. Here, the analytical results are validated using harmonic excitation, which differs from the transient excitation conditions considered in Supplementary Note 3, namely initial displacement, initial velocity, and external step force.

A continuous harmonic excitation at a selected modal frequency preferentially excites the corresponding spatial mode. The forcing location also influences modal participation. For every mode considered here, including both bulk and edge modes, the first cell has a nonzero modal amplitude. A sinusoidal force applied to the first cell can therefore couple to each selected mode.

For a targeted bulk mode, the resulting wave propagates from the first cell toward the opposite boundary and gradually develops a standing-wave pattern. For an edge mode, the first cell has the dominant modal amplitude, which decays exponentially into the chain. Excitation at the first cell therefore couples effectively to the localized mode.

Because the harmonic force was applied continuously, its amplitude was chosen to be substantially smaller than that used for the transient external-force condition in Supplementary Note 3. Specifically, the 32 N external step force was replaced by the translational sinusoidal force

$$F(t) = (0.001\,\text{N}) \sin\left(2\pi f_{\text{force}} t\right). \tag{S88}$$

The small forcing amplitude ensured that the response remained close to the linear regime in which the modal analysis was derived.

Figure S4 shows the responses obtained by exciting configurations i–iv at selected modal frequencies. The fifth mode was selected for configuration i, the tenth mode for configuration ii, and the twentieth mode for configurations iii and iv. Figure S4a presents the time–space responses obtained by applying the sinusoidal force until a stable standing-wave pattern developed. Configuration ii required a comparatively longer excitation duration because its selected mode is a relatively high-order bulk mode.

After the standing-wave pattern had developed, the spatial profile was extracted at the instant of maximum response amplitude, as shown in Figure S4b. The resulting profile in the $\widetilde{\boldsymbol{q}}$ coordinate was transformed according to Equation (6) of the main text into the $\boldsymbol{\phi}$ coordinate. The transformed profile was then normalized by its maximum absolute value, as shown in Figure S4c.

For configurations iii and iv, the spatial profiles remain nearly unchanged after the coordinate transformation. This occurs because their edge-mode amplitudes decay as successive powers of $r_\mathrm{e}$, making the cumulative contribution from the subsequent cells negligible. This procedure was used to generate the simulated mode profiles shown in Figure 3b of the main text.

Supplementary Note 5: Detailed derivation of the edge mode condition

The condition for the existence of an edge mode derived in the main text is

$$-1 < \frac{\gamma}{\gamma - \mathfrak{g}} < 1. \tag{S89}$$

Since

$$\mathfrak{g} = \vec{g}^T \hat{\mathbf{M}} \vec{g} > 0 \tag{S90}$$

for any configuration, Equation (S89) can be written as

$$\gamma^2 < (\gamma - \mathfrak{g})^2, \tag{S91}$$

which reduces to

$$2\gamma < \mathfrak{g}. \tag{S92}$$

Using the definitions of $\gamma$ and $\mathfrak{g}$, this condition becomes

$$2\left(m^{\mathrm{L}} G_1 g_1 + m^{\mathrm{R}} G_4 g_4\right) < m^{\mathrm{L}} g_1^2 + m^{\mathrm{R}} g_4^2. \tag{S93}$$

Here, the rotational inertia terms do not appear because the corresponding components of $\vec{g}$ vanish.

The left and right bars are assumed to have the same uniform linear mass density. Their masses are therefore proportional to their lengths:

$$\frac{m^{\mathrm{R}}}{m^{\mathrm{L}}} = \frac{R}{L} = \Lambda. \tag{S94}$$

For the $L$-type configuration, $q = \theta^{\mathrm{R}}$ and $\Lambda < 1$. From Equations. (S33) and (S34), the relevant Jacobian components are

$$G_1 = -R \frac{\sin\theta^{\mathrm{R}}}{\sin\theta^{\mathrm{L}}} \cos\theta^{\mathrm{L}} - 2R\cos\theta^{\mathrm{R}}, \tag{S95}$$

$$G_4 = -R\cos\theta^{\mathrm{R}}, \tag{S96}$$

$$g_1 = g_4 = -2R \frac{\sin\theta^{\mathrm{R}}}{\sin\theta^{\mathrm{L}}} \cos\theta^{\mathrm{L}} - 2R\cos\theta^{\mathrm{R}}. \tag{S97}$$

Substituting these expressions and $m^{\mathrm{R}} = \Lambda m^{\mathrm{L}}$ into Equation (S93) gives

$$\mathfrak{g}_L - 2\gamma_L = 4m^{\mathrm{L}} R^2 \left(\frac{\sin\theta^{\mathrm{R}}}{\sin\theta^{\mathrm{L}}} \cos\theta^{\mathrm{L}} + \cos\theta^{\mathrm{R}}\right)\left(\Lambda \frac{\sin\theta^{\mathrm{R}}}{\sin\theta^{\mathrm{L}}} \cos\theta^{\mathrm{L}} - \cos\theta^{\mathrm{R}}\right). \tag{S98}$$

The geometric constraint of Equation (S9) gives

$$\cos\theta^{\mathrm{L}} = \Lambda\cos\theta^{\mathrm{R}}. \tag{S99}$$

Using this relation, Equation (S98) can be rewritten as

$$\mathfrak{g}_L - 2\gamma_L = 4m^{\mathrm{L}} R^2 \cos^2\theta^{\mathrm{R}} \left(1 + \frac{\tan\theta^{\mathrm{R}}}{\tan\theta^{\mathrm{L}}}\right)\left(\Lambda \frac{\tan\theta^{\mathrm{R}}}{\tan\theta^{\mathrm{L}}} - 1\right). \tag{S100}$$

For the admissible $L$-type geometric branches considered here,

$$-1 < \frac{\tan\theta^{\mathrm{R}}}{\tan\theta^{\mathrm{L}}} < \frac{1}{\Lambda}. \tag{S101}$$

Consequently,

$$1 + \frac{\tan\theta^{\mathrm{R}}}{\tan\theta^{\mathrm{L}}} > 0, \qquad \Lambda \frac{\tan\theta^{\mathrm{R}}}{\tan\theta^{\mathrm{L}}} - 1 < 0, \tag{S102}$$

and therefore

$$2\gamma_L > \mathfrak{g}_L. \tag{S103}$$

Thus, the edge-mode condition in Equation (S92) is not satisfied for the $L$-type configuration.

For the $R$-type configuration, $q = \theta^{\mathrm{L}}$ and $\Lambda > 1$. From Equations. (S36) and (S37), the relevant Jacobian components are

$$G_1 = -2L\frac{\sin\theta^{\mathrm{L}}}{\sin\theta^{\mathrm{R}}}\cos\theta^{\mathrm{R}} - L\cos\theta^{\mathrm{L}}, \tag{S104}$$

$$G_4 = -L\frac{\sin\theta^{\mathrm{L}}}{\sin\theta^{\mathrm{R}}}\cos\theta^{\mathrm{R}}, \tag{S105}$$

$$g_1 = g_4 = -2L\frac{\sin\theta^{\mathrm{L}}}{\sin\theta^{\mathrm{R}}}\cos\theta^{\mathrm{R}} - 2L\cos\theta^{\mathrm{L}}. \tag{S106}$$

Substituting these expressions and $m^{\mathrm{R}} = \Lambda m^{\mathrm{L}}$ into Equation (S93) yields

$$\mathfrak{g}_R - 2\gamma_R = 4m^{\mathrm{L}}L^2\left(\frac{\sin\theta^{\mathrm{L}}}{\sin\theta^{\mathrm{R}}}\cos\theta^{\mathrm{R}} + \cos\theta^{\mathrm{L}}\right)\left(\Lambda\cos\theta^{\mathrm{L}} - \frac{\sin\theta^{\mathrm{L}}}{\sin\theta^{\mathrm{R}}}\cos\theta^{\mathrm{R}}\right). \tag{S107}$$

Using the geometric constraint, Equation (S107) becomes

$$\mathfrak{g}_R - 2\gamma_R = 4m^{\mathrm{L}}L^2\cos^2\theta^{\mathrm{L}}\left(1 + \frac{\tan\theta^{\mathrm{L}}}{\tan\theta^{\mathrm{R}}}\right)\left(\Lambda - \frac{\tan\theta^{\mathrm{L}}}{\tan\theta^{\mathrm{R}}}\right). \tag{S108}$$

For the admissible $R$-type geometric branches,

$$-1 < \frac{\tan\theta^{\mathrm{L}}}{\tan\theta^{\mathrm{R}}} < \Lambda. \tag{S109}$$

Thus,

$$1 + \frac{\tan\theta^{\mathrm{L}}}{\tan\theta^{\mathrm{R}}} > 0, \qquad \Lambda - \frac{\tan\theta^{\mathrm{L}}}{\tan\theta^{\mathrm{R}}} > 0, \tag{S110}$$

which gives

$$2\gamma_R < \mathfrak{g}_R. \tag{S111}$$

Therefore, the edge-mode condition is always satisfied for the $R$-type configuration.

Hence, under the assumption of a uniform linear mass density, the edge mode exists for the admissible $R$-type asymmetric configurations but not for the corresponding $L$-type configurations.

Supplementary Note 6: Detailed derivation of the singular edge mode condition

The singular edge mode corresponds to the limiting condition $r = 0$. From the decay factor of the $R$-type configuration,

$$r = \frac{\gamma_R}{\gamma_R - \mathfrak{g}} = 0 \qquad \Longrightarrow \qquad \gamma_R = 0. \tag{S112}$$

The coefficient $\gamma_R$ is defined as

$$\gamma_R = \vec{G}_R^T \hat{\mathbf{M}} \vec{g}_R = G_1 m^{\mathrm{L}} g_1 + G_4 m^{\mathrm{R}} g_4. \tag{S113}$$

Substituting the corresponding Jacobian components from Equations. (S36) and (S37) gives

$$\begin{aligned}\gamma_R = &\left(-\frac{2L\sin\theta^{\mathrm{L}}}{\sin\theta^{\mathrm{R}}}\cos\theta^{\mathrm{R}} - L\cos\theta^{\mathrm{L}}\right) m^{\mathrm{L}} \left(-\frac{2L\sin\theta^{\mathrm{L}}}{\sin\theta^{\mathrm{R}}}\cos\theta^{\mathrm{R}} - 2L\cos\theta^{\mathrm{L}}\right) \\ &+ \left(-\frac{L\sin\theta^{\mathrm{L}}}{\sin\theta^{\mathrm{R}}}\cos\theta^{\mathrm{R}}\right) m^{\mathrm{R}} \left(-\frac{2L\sin\theta^{\mathrm{L}}}{\sin\theta^{\mathrm{R}}}\cos\theta^{\mathrm{R}} - 2L\cos\theta^{\mathrm{L}}\right).\end{aligned} \tag{S114}$$

Assuming that the left and right bars have the same uniform linear mass density, their masses are proportional to their lengths:

$$\frac{m^{\mathrm{R}}}{m^{\mathrm{L}}} = \frac{R}{L} = \Lambda. \tag{S115}$$

Using Equation (S115) and collecting the common factors, the condition $\gamma_{\mathrm{R}} = 0$ can be written as

$$\gamma_{\mathrm{R}} = 2m^{\mathrm{L}}L^2 \left[(\Lambda + 2)\frac{\sin\theta^{\mathrm{L}}}{\sin\theta^{\mathrm{R}}}\cos\theta^{\mathrm{R}} + \cos\theta^{\mathrm{L}}\right]\left[\frac{\sin\theta^{\mathrm{L}}}{\sin\theta^{\mathrm{R}}}\cos\theta^{\mathrm{R}} + \cos\theta^{\mathrm{L}}\right] = 0. \tag{S116}$$

The vanishing of the second factor in Equation (S116) would require

$$\frac{\tan\theta^{\mathrm{R}}}{\tan\theta^{\mathrm{L}}} = -1, \tag{S117}$$

which is excluded for the admissible asymmetric configurations considered here. Therefore, the first factor must vanish, yielding the closed-form geometric condition for the singular edge mode:

$$\frac{\tan\theta^{\mathrm{R}}}{\tan\theta^{\mathrm{L}}} = -(\Lambda + 2). \tag{S118}$$

Supplementary Note 7: Additional information on the singular edge mode

Figure S5 presents simulation results for the singular configuration, configuration v, under two additional input conditions distinct from that considered in the main text. The initial-displacement results shown in Figure S5a–d are nearly identical to those shown in Figure 4b–e of the main text. However, imposing a relatively large initial displacement of 15° on the first cell induces nonlinear effects, causing slight leakage into the bulk response. This results in a small frequency shift, as shown in Figure S5b, and a slight reduction in the IPR, as shown in Figure S5c. Nevertheless, the peak normalized energy in Figure S5d decreases to approximately $10^{-4}$ at the distal end of the chain. Therefore, even under the initial-displacement condition, the response can still be characterized as extremely localized.

The external-force results shown in Figure S5e–h exhibit a particularly interesting response. The edge mode remains strongly excited, as under the other input conditions; however, its influence extends more noticeably into the subsequent cells. At first glance, this behavior appears counterintuitive because the singular edge mode is expected to remain spatially isolated from the rest of the chain. To clarify this discrepancy, we examine the Fourier domain result in Figure S5f. The observed edge-mode frequency is considerably lower than the analytical eigenfrequency obtained from the modal analysis, which closely matches the edge-mode frequency observed under the other input conditions. This difference indicates a pronounced amplitude-dependent frequency shift.

In the main-text simulation under the initial-velocity condition, the edge-mode frequency shifts by approximately 5.8%. This shift is consistent with an amplitude-dependent behavior under finite-amplitude excitation. Under the external-force condition considered here, a substantially larger shift of approximately 33.3% occurs, causing the shifted edge-mode frequency to enter the bulk-mode frequency range. Under external-force excitation, the dominant oscillation frequency of the singular edge mode is not determined solely by its linear eigenfrequency but is collectively modified at finite amplitudes by the configuration-dependent mass matrix, velocity-dependent inertial terms, and nonlinear restoring force.

As shown in Figure S5f, the shifted edge-mode frequency overlaps with the bulk-mode frequency range. This spectral overlap activates bulk-mode components and thereby disrupts the complete spatial isolation of the singular edge mode. Such bulk-mode activation is also reflected in the reduced IPR values shown in Figure S5g. The peak normalized energy in Figure S5h still indicates substantial localization; however, the localization is weaker than under the other excitation conditions, with the energy at the distal end remaining at approximately the $10^{-3}$ level.

To further examine the frequency-shift behavior, additional simulations were performed by varying the excitation amplitude. Because the initial-displacement condition has practical limitations on the applicable amplitude range, this analysis was conducted only for the initial-velocity and external-force conditions. As shown in Figure S6a–d, the frequency of the nearly horizontal edge-mode feature progressively decreases as the initial-velocity amplitude increases. At the smallest excitation amplitude, the extracted frequency closely agrees with the edge-mode frequency predicted by the linear modal analysis. In particular, the bulk response, which is barely visible at small excitation amplitudes, becomes increasingly pronounced as the amplitude increases. At sufficiently large amplitudes, the shifted edge-mode frequency approaches the upper-frequency region of the bulk band.

A similar trend is observed under the external-force condition, as shown in Figure S6e–h. At the smallest force amplitude, the extracted edge-mode frequency again closely agrees with the prediction of the linear modal analysis. As the force amplitude increases, the edge-mode frequency shifts downward. Compared with the initial-velocity condition, the external-force condition induces a more pronounced frequency shift, and in Figure S6h, the shifted edge-mode feature substantially overlaps with the bulk band. This spectral overlap produces a mixed edge- and bulk-mode response, explaining why the spatial attenuation under external-force excitation is weaker than that under the initial-velocity and initial-displacement conditions.

Modeling this frequency shift requires consideration not only of stiffness nonlinearity but also of the configuration- and velocity-dependent inertial nonlinearities of the system. A more comprehensive analysis should also account for additional factors, including different boundary conditions, nonuniform arrangements, and possible topological extensions. Future work will therefore focus on developing a mathematical model for the amplitude-dependent frequency shift and investigating strategies to improve spatial attenuation under external-force excitation. In particular, because the system exhibits a softening trend in which the oscillation frequency decreases with increasing amplitude, its nonlinear continuation and stability properties warrant further investigation. Such an analysis would provide insight into both the underlying wave dynamics and the design of practical devices based on this localization mechanism.

Supplementary Note 8: Additional information on the experiments

***Fourier-domain analysis of the experiments*** — To further examine the experimentally observed wave characteristics, a two-dimensional Fourier transform was applied to the measured time–space displacement fields (Figure S7). Owing to the relatively small number of cells in the experimental chain, the resulting spectra have limited wavenumber resolution; nevertheless, clear configuration-dependent trends can be identified. Configurations i and ii exhibit pronounced bulk bands, consistent with their strong wave-transmission behavior. In configuration iii, a weak signature of a relatively flat edge-mode branch begins to emerge, although it is not clearly resolved from the coexisting bulk response. In contrast, the bulk contribution becomes substantially weaker for configurations iv and v, while distinct edge-mode responses appear around 15.6 Hz and 14.9 Hz, respectively. The normalized spectral intensity of configuration v is slightly lower, which is attributed to the stronger dissipation observed in this configuration.

***Details of the unit cells*** — Accurately reproducing the prescribed asymmetric-linkage geometry while minimizing fabrication and assembly tolerances was essential for the experiments. To realize the designed geometry as precisely as possible, the unit cells were fabricated using the Extra Fine profile in Bambu Studio, with a layer height of 0.08 mm. A customized three-dimensional geometry was adopted to prevent interference between adjacent unit cells while minimizing the intercell spacing. The reduced spacing shortened the overall span of the experimental chain and thereby helped limit shaft deflection. The links were fabricated with a thickness of at least 6 mm to suppress unintended bending and maintain the intended rigid-link approximation.

A substantial source of assembly tolerance was the installation of the torsional springs at the linkage joints. To minimize this uncertainty, snap-fit mounting grooves were carefully dimensioned and fabricated to match the geometries of the selected torsional springs. Four types of torsional springs from MISUMI were used to provide the restoring stiffness of the linkage units: C-UA90L6-4-1, C-UA90R6-4-1, C-UA180L6-4-1, and C-UA180R6-4-1. According to the manufacturer specifications, C-UA90L6-4-1 and C-UA90R6-4-1 have a torsional spring constant of 1.4045 N mm/deg, whereas C-UA180L6-4-1 and C-UA180R6-4-1 have a torsional spring constant of 1.476 N mm/deg. The left- and right-handed spring variants were selected according to the orientation and equilibrium configuration of each unit cell.

To reduce rotational friction and assembly-induced variability, all revolute joints were equipped with economy-type deep-groove ball bearings from MISUMI (C-E682ZZ). The bearings also helped maintain joint alignment and minimize unintended resistance associated with direct contact between the printed components and the joint shafts.

***Details of the experimental setup*** — In one-dimensional dynamic experiments such as those considered here, minimizing friction between the moving components and the supporting guide is essential. The experimental system was constructed using straight shafts (MISUMI, E-SSFJ6-800) and straight linear bushings (MISUMI, E-LB6UU), with each bushing incorporated into a sliding base supporting one unit cell. Among the available linear bushings, those exhibiting the smoothest motion and lowest apparent friction were selected manually. The selected bushings and shafts were lubricated before the experiments to further reduce sliding resistance.

The sliding base, including the linear bushing designed for a 6 mm-diameter shaft, had a mass of approximately 7.9 g. Minimizing this additional moving mass was desirable for reproducing the intended dynamics of the linkage chain without introducing excessive inertia from the supporting components. Although a thinner shaft could further reduce the mass of the associated bushing and sliding base, the relatively long unsupported span of the shaft would make it more susceptible to bending and external vibrations. A shaft diameter of 6 mm was therefore selected as a practical compromise between minimizing the moving mass and maintaining sufficient structural rigidity and vibration resistance.

An initial version of the setup employed two parallel shafts to prevent rigid-body rotation of the sliding bases about the shaft direction. However, contact with two guide shafts introduced additional friction and increased sensitivity to small alignment errors. The setup was therefore modified to use a single shaft for each sliding base. To reproduce the fixed boundary condition while preventing rotation about the shaft axis, the final cell was rigidly constrained against both translation and axial rotation. Because the remaining cells were connected through hinges whose rotational axes were orthogonal to the shaft direction, this terminal constraint effectively suppressed rigid-body axial rotation throughout the connected chain while preserving the intended in-plane linkage motion.

The shafts were mounted on aluminum frames fixed to an optical table. Their alignment and horizontality were carefully adjusted using a spirit level to avoid gravitational drift caused by shaft inclination. The relative alignment of the shafts was also checked to minimize unintended resistance or geometric distortion during sliding.

For video recording, an additional aluminum support frame was installed above the experimental system. A dedicated camera bracket was attached to this frame to maintain a fixed camera position and viewing direction throughout the repeated experiments. The motion was recorded using an iPhone 14 Pro (Apple Inc., Cupertino, CA, USA) positioned approximately 0.8 m above the midpoint of the shaft array. The rear camera was operated in the built-in slow-motion mode at a resolution of $1920 \times 1080$ pixels and an acquisition rate of 240 frames $s^{-1}$, corresponding to a temporal interval of approximately 4.17 ms between consecutive frames. All tracking, synchronization, and subsequent data processing were performed using the original acquisition rate of 240 frames $s^{-1}$, rather than the reduced playback speed of the slow-motion video.

***Details of the excitation method*** — As discussed for the numerical simulations, three types of transient input conditions were considered: initial-displacement, initial-velocity, and external-force. Implementing the latter two conditions experimentally would require additional devices. For example, an external-force input could be generated by sliding a ball into the structure or by applying an impact with a hammer, whereas an initial-velocity input could be imposed using a motorized actuator that releases the constrained cell at a prescribed time. Although both approaches were tested, it was difficult to achieve the desired input profile and timing while also separating the excitation device from the system immediately after excitation. Residual contact with the device could additionally influence the subsequent wave propagation.

We therefore adopted an initial-displacement excitation, which provided the simplest, most direct, and most repeatable implementation. For each configuration, the first unit cell was displaced by 30° from its equilibrium position and firmly constrained using a thin metal wire arranged as a quick-release loop. The excitation was initiated by cutting one of the loops, causing the constraint to release rapidly. This method enabled the prescribed initial displacement to be set consistently and allowed the excitation mechanism to be separated from the system immediately after release, thereby minimizing its unintended influence on the subsequent wave propagation.

***Experimental limitations*** — The experimental system inevitably contains dissipative effects that are not included in the ideal numerical model. These include material damping in the PLA components, intrinsic damping and possible hysteresis in the torsional springs, and sliding friction between the linear bushings and shafts. Together, these effects attenuate the propagating response and contribute to quantitative differences between the experimental and numerical results.

The experimental chain was therefore constructed using 11 unit cells. When chains containing 15 or more cells were considered, a larger excitation amplitude exceeding the present initial displacement of 30° was required for the propagating wave to overcome dissipation and remain measurable at the distal end. Such a large input would drive the linkage deeply into the nonlinear regime, making direct comparison with the linear modal predictions and the moderate-amplitude numerical results less reliable. The 11-cell configuration was consequently selected as a practical compromise between observing spatial wave propagation and limiting the influence of strong nonlinearity.

Changes in gravitational potential energy also arise because the centers of mass of the links move vertically during the linkage motion. However, under the excitation conditions considered here, the combined kinetic and torsional elastic potential energies were more than one order of magnitude larger than the change in gravitational potential energy. The gravitational contribution was therefore neglected in the cell-wise energy calculation for simplicity. More ideal experimental conditions could be achieved by using an air-bearing guide system to further reduce friction and by fabricating the unit cells through precision CNC machining of metal components to reduce material damping and manufacturing tolerances. These alternatives would, however, substantially increase the complexity and cost of the experimental setup. Despite the remaining nonideal effects, the present setup provides sufficiently clear and repeatable measurements to identify the configuration-dependent transition between bulk transmission and strong boundary localization.

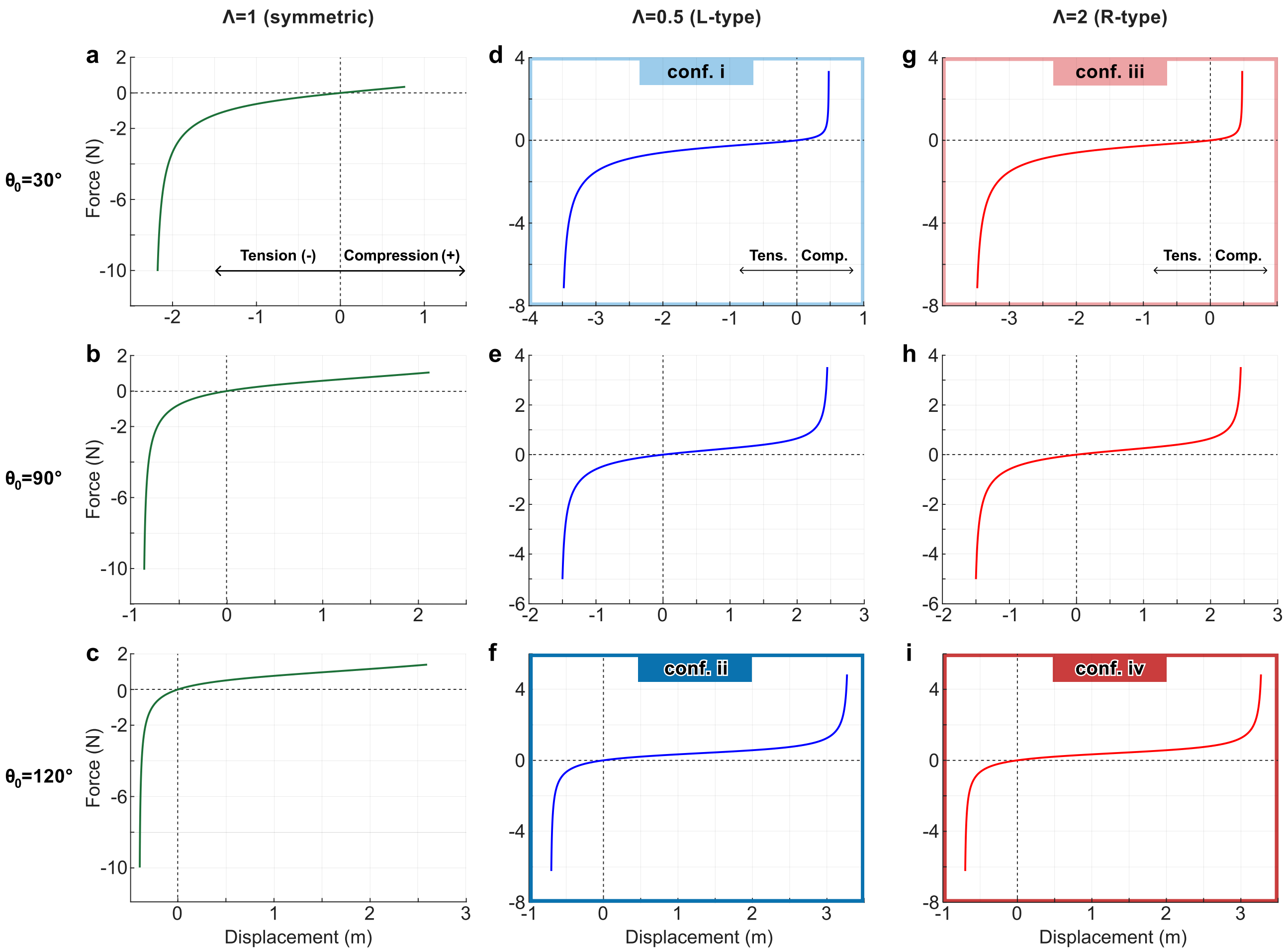

**Figure S1. Static force–displacement responses across symmetric and asymmetric linkage configurations.** (a–c) Results for the symmetric linkage unit cell (Λ = 1). (d–f) Results for the asymmetric *L*-type linkage unit cells (Λ = 0.5), including configurations i and ii. (g–i) Results for the asymmetric *R*-type linkage unit cells (Λ = 2), including configurations iii and iv.

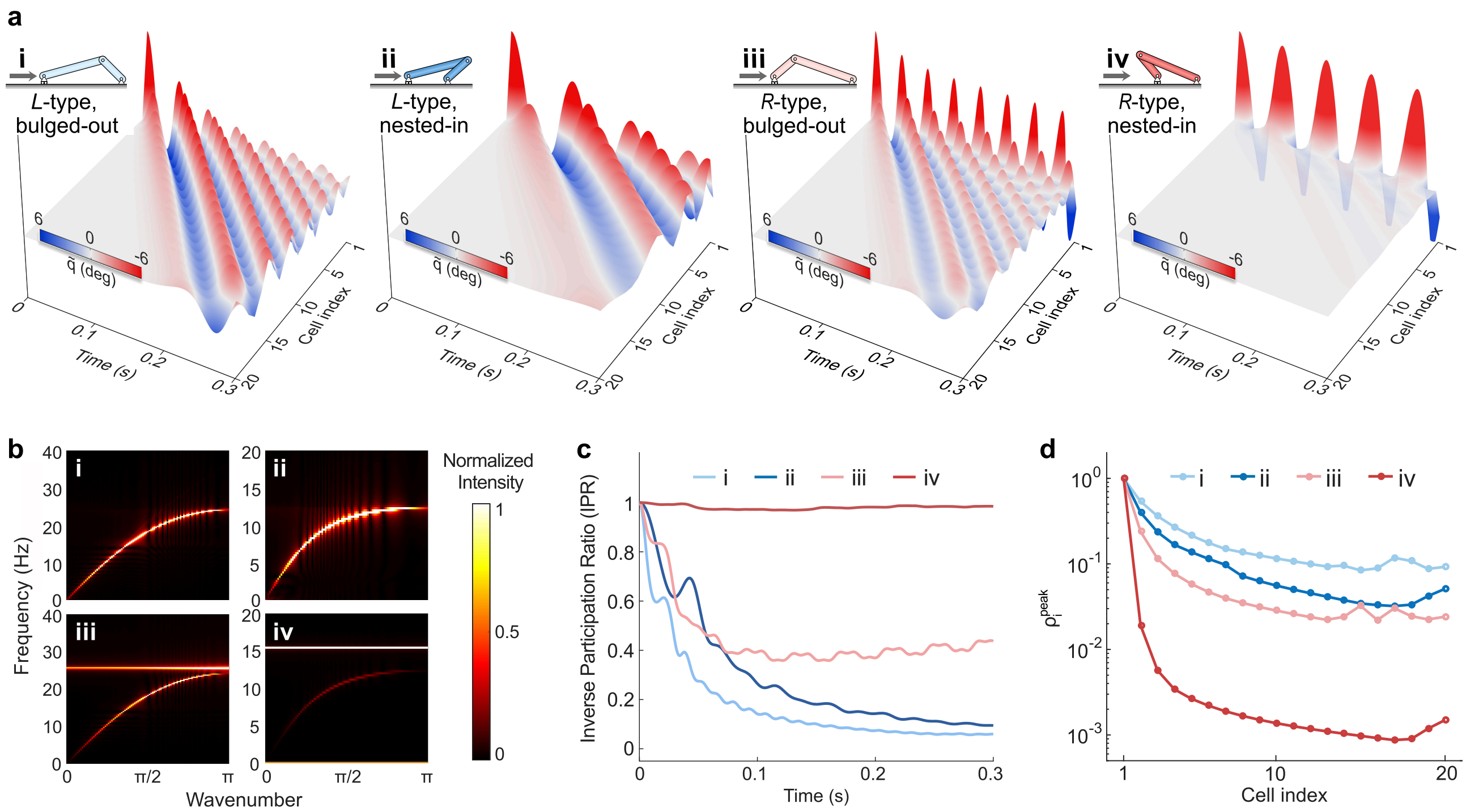

**Figure S2. Initial-displacement excitation produces configuration-dependent wave transmission and localization.** (a) Numerical time–space responses of the generalized coordinate for configurations i–iv. (b) Corresponding two-dimensional Fourier spectra, computed using a chain of 200 cells to improve wavenumber resolution. (c) Time evolution of the inverse participation ratio (IPR). (d) Cell-wise normalized peak energy on a semilogarithmic scale.

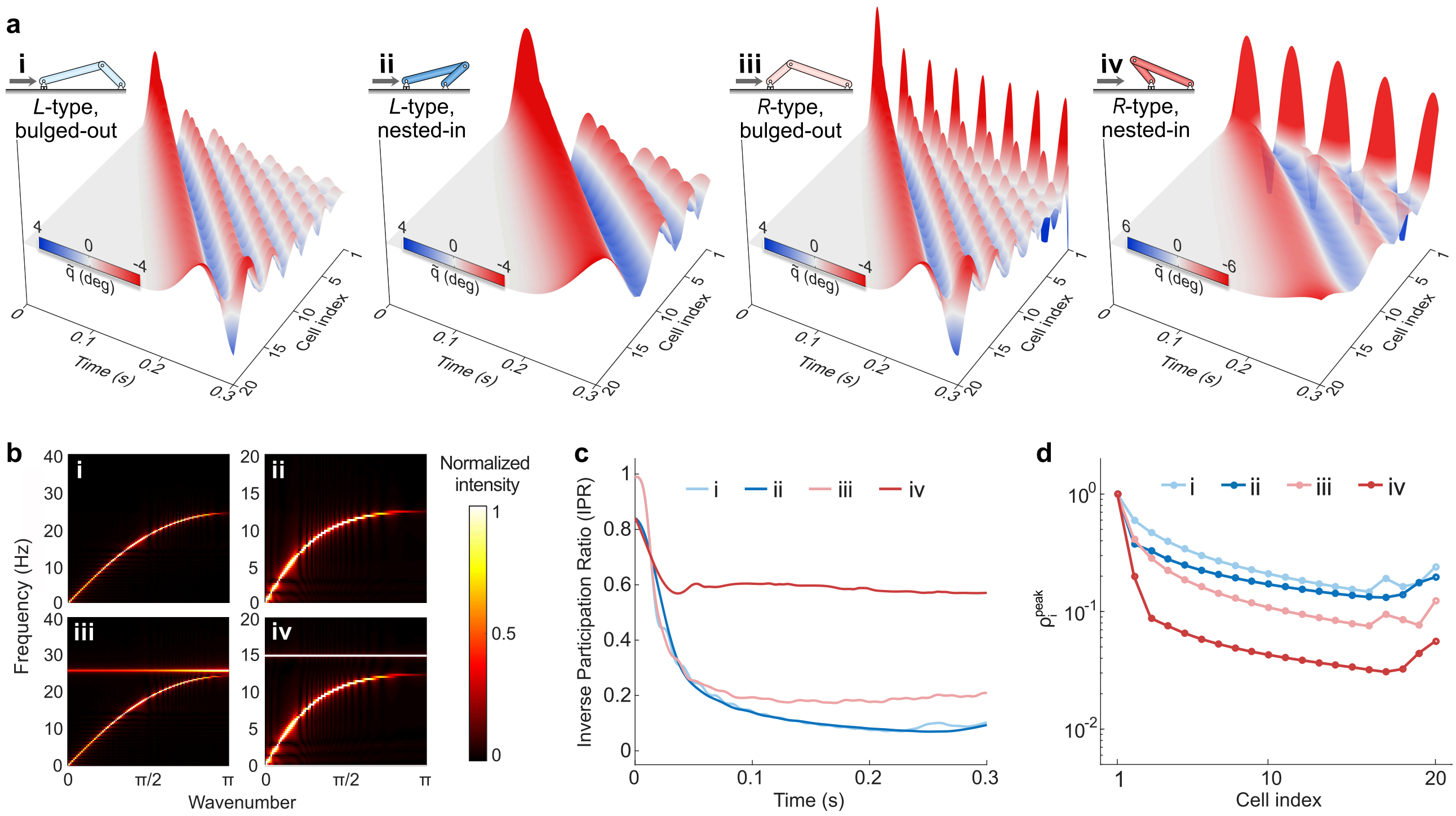

**Figure S3. External-force excitation produces configuration-dependent wave transmission and localization.** (a) Numerical time–space responses of the generalized coordinate for configurations i–iv. (b) Corresponding two-dimensional Fourier spectra, computed using a chain of 200 cells to improve wavenumber resolution. (c) Time evolution of the inverse participation ratio (IPR). (d) Cell-wise normalized peak energy on a semilogarithmic scale.

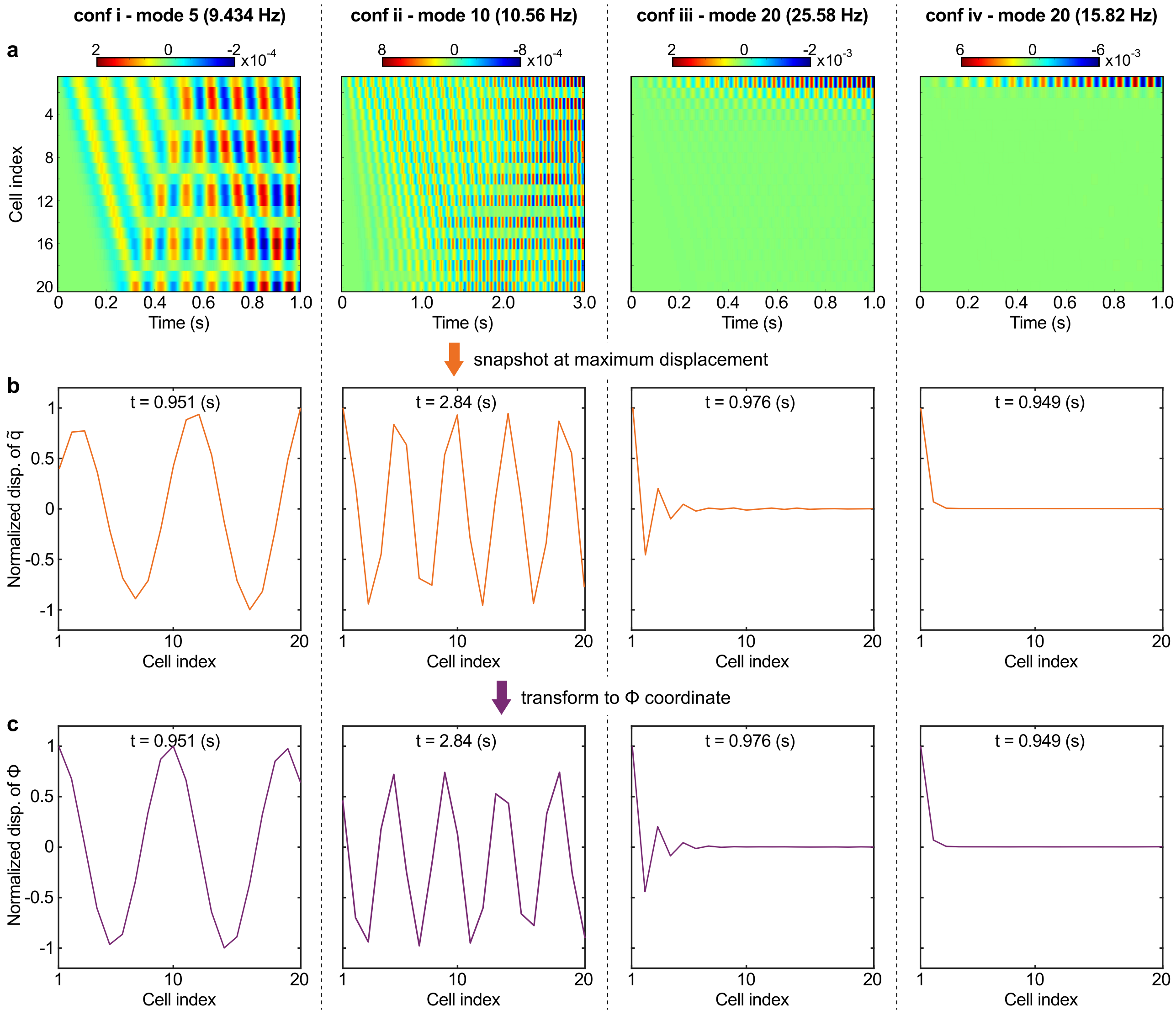

**Figure S4. Procedure for generating the sinusoidal-input simulation profiles shown in Figure 3b of the main text.** (a) Numerical responses of configurations i–iv, shown as time–space contour plots of the generalized coordinate. (b) Normalized displacement profiles in the $\tilde{q}$ coordinate, extracted at the instant of maximum response amplitude after a stable standing-wave pattern had developed. (c) Corresponding normalized profiles transformed into the $\boldsymbol{\phi}$ coordinate.

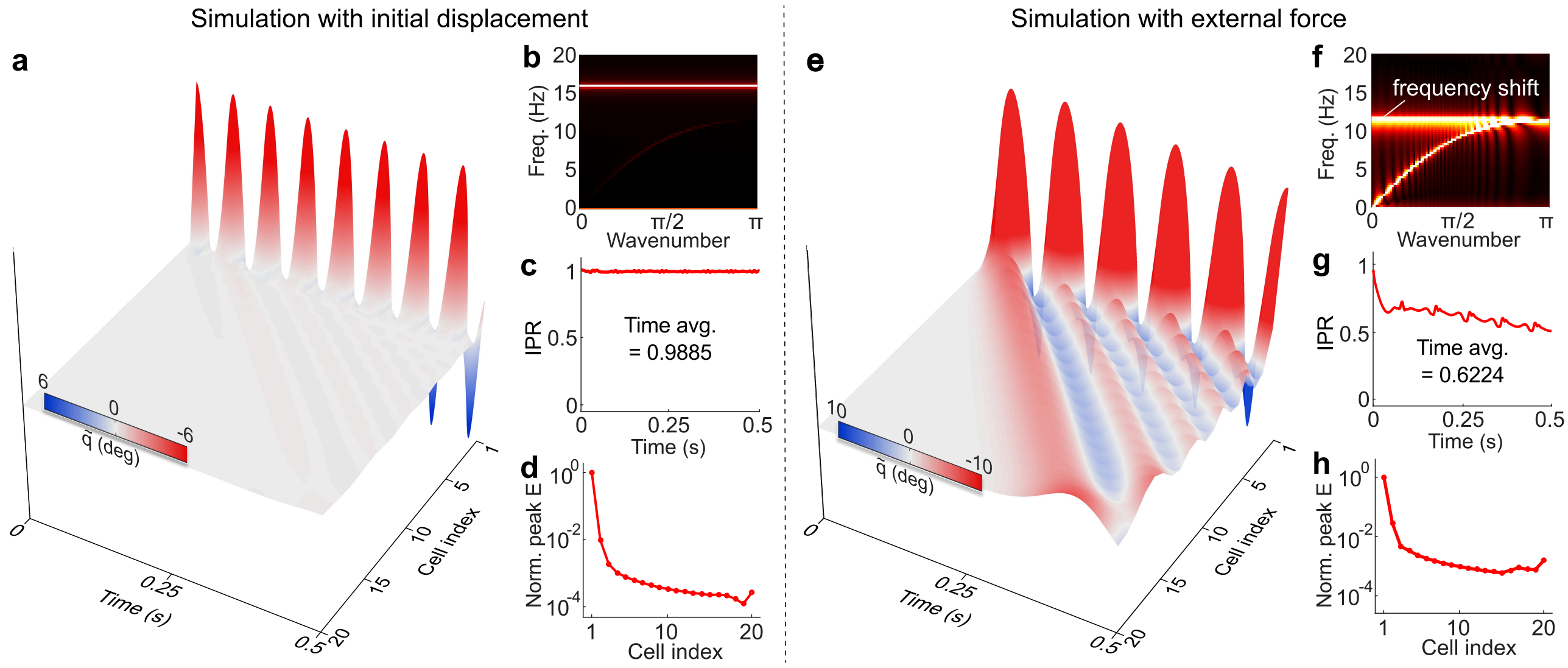

**Figure S5. Configuration v remains strongly localized under different excitation conditions.** (a–d) Numerical response of configuration v under initial-displacement excitation, together with the corresponding Fourier spectrum, IPR, and cell-wise normalized peak energy. (e–h) Numerical response of configuration v under external-force excitation, together with the corresponding Fourier spectrum, IPR, and cell-wise normalized peak energy. For visualization, each panel is normalized independently to enhance the visibility of the edge-mode response.

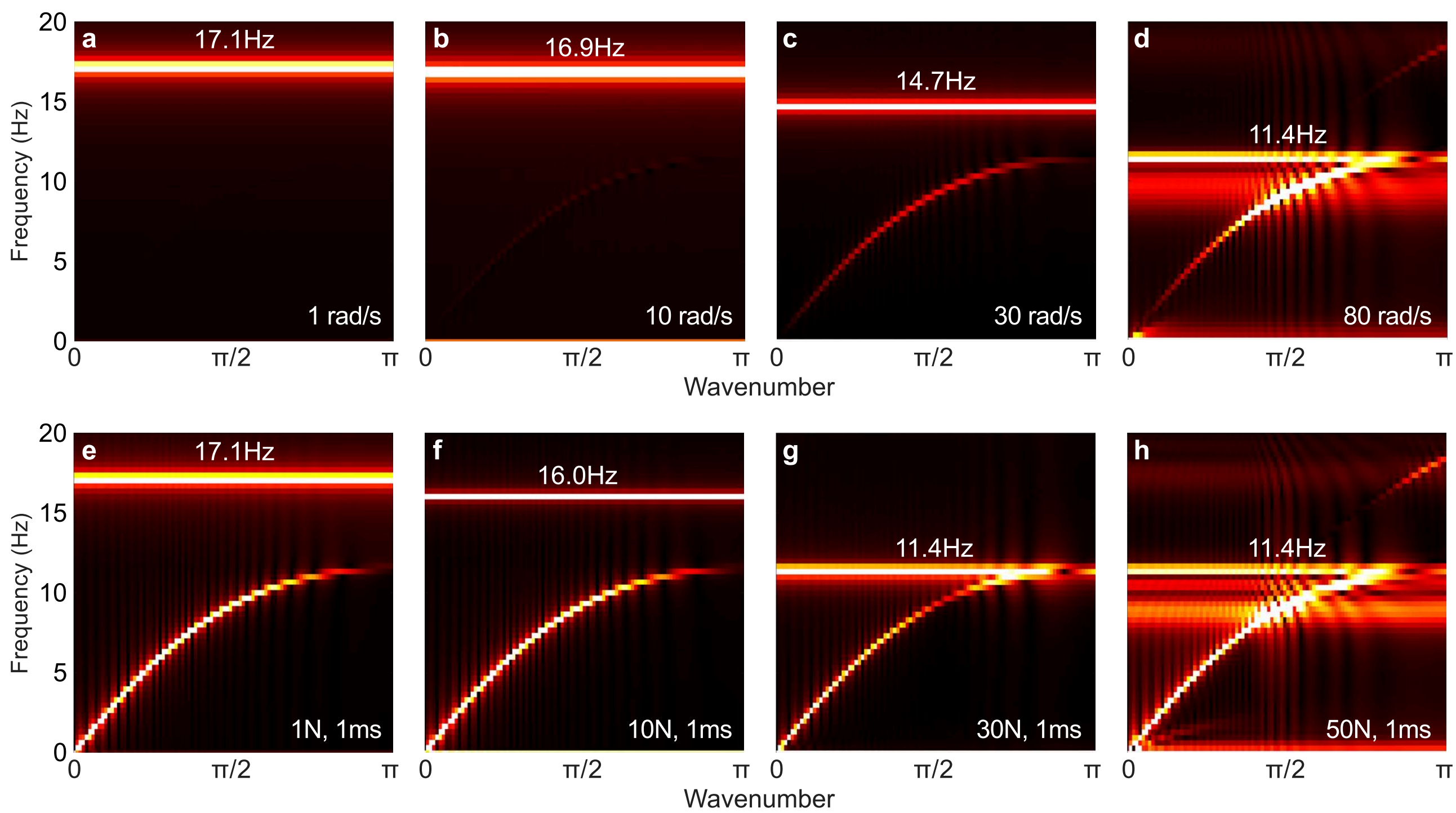

**Figure S6. The response frequency of configuration v shifts with excitation amplitude.** (a–d) Numerical responses under initial-velocity excitation for different velocity amplitudes. (e–h) Numerical responses under external-force excitation for different force amplitudes. Because the input amplitude differs among cases, each response is independently normalized to clearly visualize the corresponding edge mode.

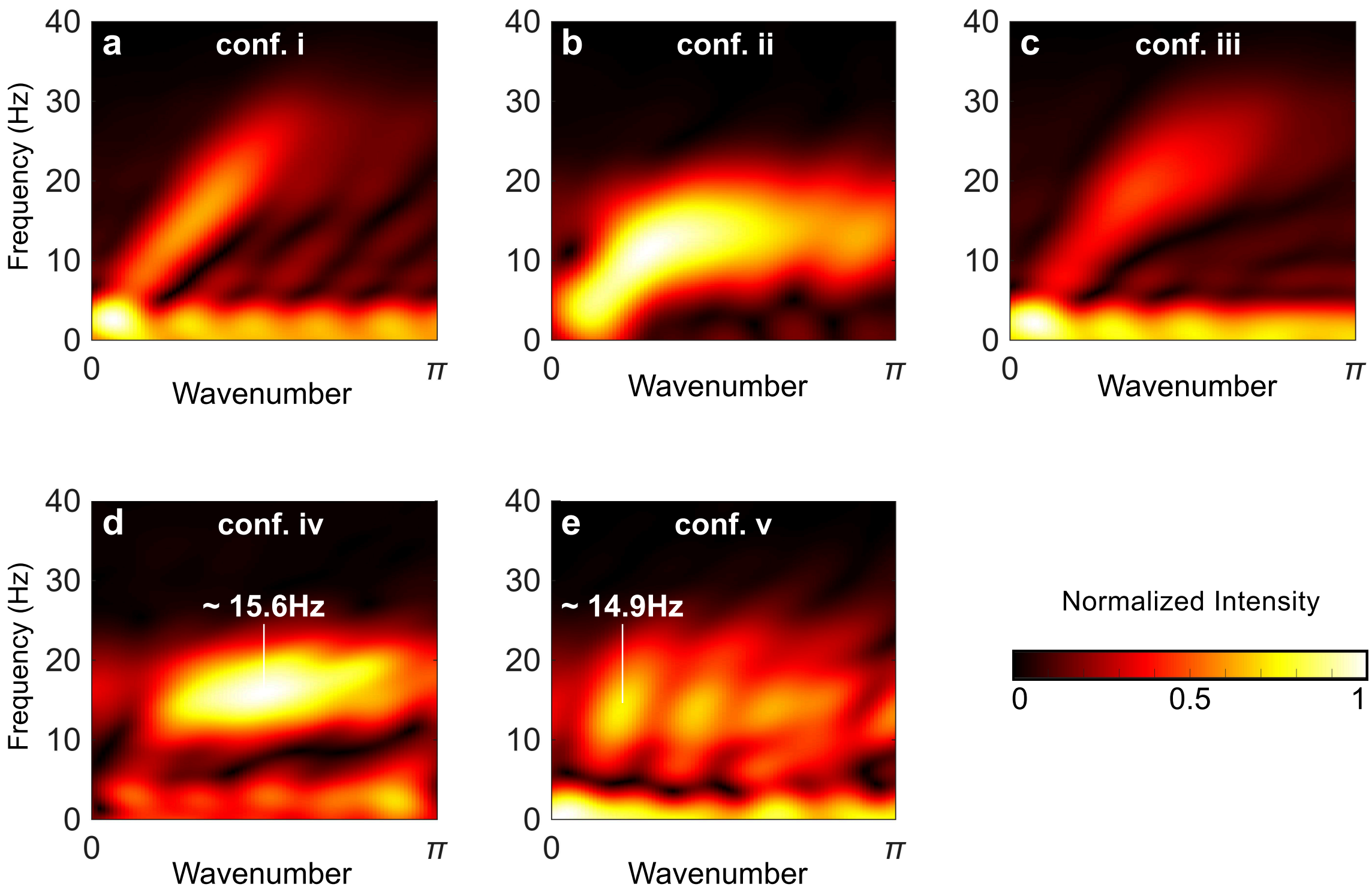

**Figure S7. Experimental Fourier spectra support the configuration-dependent emergence of edge-mode responses.** Two-dimensional Fourier spectra obtained from the experimentally measured time–space displacement fields for configurations i–v [(a)–(e), respectively]. Configurations i and ii exhibit pronounced bulk-band responses, whereas configuration iii shows a weak edge-mode signature superposed on a dominant bulk response. In configurations iv and v, the bulk contribution is strongly suppressed and distinct edge-mode responses appear near 15.6 Hz and 14.9 Hz, respectively. Spectral intensity is normalized independently for each configuration.